\documentclass[USenglish, 12pt, fleqn]{article}
\usepackage[T1]{fontenc}
\usepackage{textcomp}
\usepackage[USenglish=usenglishmax]{hyphsubst}
\usepackage{babel}
\usepackage{ragged2e}
\usepackage{amsmath,amssymb,amsfonts}
\usepackage{graphicx}
\usepackage[xcdraw]{xcolor}
\usepackage[margin=1in,footskip=0.25in]{geometry}
\usepackage[numbers,sort&compress]{natbib}
\usepackage{array}
\usepackage[margin=20pt,font=small,labelfont=bf]{caption}
\usepackage[marginal]{footmisc}
\usepackage{tikz}
\usetikzlibrary{decorations.pathreplacing,decorations.pathmorphing,fit,quotes}
\usepackage[colorlinks=true]{hyperref}
\usepackage{url}
\usepackage{subcaption}
\usepackage{authblk}
\usepackage{yquant}
\usetikzlibrary{fit,quotes}
\usepackage[mathletters]{ucs} \usepackage[utf8x]{inputenc}
\usepackage{mathtools}

\usepackage[frozencache=true,cachedir=.]{minted}
\definecolor{mintedbackground}{rgb}{0.95,0.95,0.95}

\newmintedfile[qasmfile]{qasmlexer.py:QASMLexer -x}{
bgcolor=mintedbackground,
fontsize=\scriptsize,
linenos=true,
 }
 
 \newmintedfile[qasmfilewithlines]{qasmlexer.py:QASMLexer -x}{
bgcolor=mintedbackground,
fontsize=\scriptsize,
linenos=true,
}

\newcommand*{\qasmline}{\mintinline[fontfamily=tt,breaklines]{qasmlexer.py:QASMLexer -x}}

\newminted[qasmblock]{qasmlexer.py:QASMLexer -x}{
bgcolor=mintedbackground,
fontsize=\scriptsize,
linenos=true,
} 

\newminted[pythonblock]{python}{
bgcolor=mintedbackground,
fontsize=\scriptsize
} 

\providecommand{\ket}[1]{\ensuremath{\left|#1 \right\rangle}}

\newcommand{\spring}[3][0mm]{
    \def \shiftamount {#1}
    \draw[decoration={aspect=0.1, segment length=1.5mm, amplitude=#3, coil},decorate] ([yshift=-\shiftamount] #2.west) -- ([yshift=-\shiftamount] #2.east) ;
    \draw [dashed] (#2.north west) -- (#2.south west);
    \draw [dashed] (#2.north east) -- (#2.south east);
}

\yquantdefinegate{repstart} {
    qubit {} x;
    qubit {} y;
    [operators/every barrier/.append style={line width=1
    .2mm, solid}]
    barrier (x, y);
    [operators/every barrier/.append style={line width=.5mm, solid, , shorten <= 2.5mm, shorten >= 2.5mm}, operator/separation=-10pt]
    barrier (x, y);
}

\yquantdefinegate{repend}{
    qubit {} x;
    qubit {} y;
    [operators/every barrier/.append style={line width=.5mm, solid}]
    barrier (x, y);
    [operators/every barrier/.append style={line width=1.2mm, solid}, operator/separation=-10pt]
    barrier (x, y);
}

\newcommand{\circledarrow}[3]{
\draw[#1,->] (#2) +(80:#3) arc(-60:240:#3);
}

\begin{document}

\title{OpenQASM\,3: A broader and deeper quantum assembly language}

\author[1]{Andrew Cross}
\author[1]{Ali Javadi-Abhari}
\author[1]{Thomas Alexander}
\author[3]{Niel de Beaudrap}
\author[1]{Lev S. Bishop}
\author[2]{Steven Heidel}
\author[2]{Colm A. Ryan}
\author[2]{Prasahnt Sivarajah}
\author[1]{John Smolin}
\author[1]{Jay M. Gambetta}
\author[1]{Blake R. Johnson}
\affil[1]{IBM Quantum, IBM T.\ J.\ Watson Research Center, Yorktown Heights, NY}
\affil[2]{AWS Center for Quantum Computing, Pasadena, CA}
\affil[3]{Dept.\ Informatics, University of Sussex, UK}

\maketitle

\begin{abstract}
Quantum assembly languages are machine-independent languages that traditionally describe quantum computation in the circuit model. Open quantum assembly language (OpenQASM 2) was proposed as an imperative programming language for quantum circuits based on earlier QASM dialects. In principle, any quantum computation could be described using OpenQASM 2, but there is a need to describe a broader set of circuits beyond the language of qubits and gates. By examining interactive use cases, we recognize two different timescales of quantum-classical interactions: real-time classical computations that must be performed within the coherence times of the qubits, and near-time computations with less stringent timing. Since the near-time domain is adequately described by existing programming frameworks, we choose in OpenQASM 3 to focus on the real-time domain, which must be more tightly coupled to the execution of quantum operations. We add support for arbitrary control flow as well as calling external classical functions. In addition, we recognize the need to describe circuits at multiple levels of specificity, and therefore we extend the language to include timing, pulse control, and gate modifiers. These new language features create a multi-level intermediate representation for circuit development and optimization, as well as control sequence implementation for calibration, characterization, and error mitigation. 
\end{abstract}

\newpage
{
  \hypersetup{linkcolor=black}
  \tableofcontents
}
\newpage

\section{Introduction}

Open quantum assembly language (OpenQASM\,2) \cite{CBSG17} was proposed as an imperative programming language for quantum circuits based on earlier QASM dialects \cite{qasm2circ,qasmtools,svore06,quale,dousti16}, and is one of the programming interfaces of IBM Quantum Services \cite{ibmqx}. In the period since OpenQASM\,2 was introduced, it has become something of a \emph{de facto} standard, allowing a number of independent tools to inter-operate using OpenQASM\,2 as the common interchange format. It also exists within the context of significant other activity in exploring and defining quantum assembly languages for practical use \cite{Amy19,Reqasm,khammassi18, jaqal20,eqasm19, smith16,xacc20}. The features of OpenQASM\,2 were strongly influenced by the needs of the time particularly with respect to the capabilities of near-term quantum hardware. We deem it time to update OpenQASM to better support the needs of the next phase of quantum system development as well as to incorporate some of the best ideas that have arisen in the other circuit description languages. One common feature of most of these machine-independent languages is that they describe quantum computation in the quantum circuit model~\cite{deutsch89,barenco95,NC00}, which includes non-unitary primitives such as teleportation~\cite{bennett1993} and the measurement-model of quantum computing~\cite{raussendorf2001, josza2005}. For practical purposes, we see advantage in a language that goes beyond the circuit model, extending the definition of a circuit to involve a significant amount of classical control flow.

We define an extended quantum circuit as a computational routine, consisting of an ordered sequence of quantum operations on quantum data --- such as gates, measurements, and resets --- and concurrent real-time classical computation. Data flows back and forth between the quantum operations and real-time classical computation, in that the classical compute can depend on measurement results, and the quantum operations may involve or be conditioned on data from the real-time classical computation. A practical quantum assembly language should be able to describe such versatile circuits.

Recognizing that OpenQASM is also a resource for describing circuits 
that characterize, validate, and debug quantum systems, we also introduce instruction semantics that 
allow control of gate scheduling in the time domain, and that provide the ability to describe the microcoded 
implementations of gates in a way that is extensible to future developments in quantum control 
and is platform agnostic. The extension considered here is intended to be backwards compatible with OpenQASM\,2, except in some uncommonly used cases where breaking changes were necessary.

As quantum applications are developing where substantial classical processing may be used along with the extended quantum circuits, we allow for a richer family of computation that incorporates interactive use of a quantum processor~(QPU). We refer to this abstraction as a quantum program. For example, the variational quantum eigensolver depends on classical computation to calculate an energy or cost function and uses classical optimization to tune circuit parameters~\cite{vqe}. By examining the broader family of interactive use cases, we recognize two different timescales of quantum-classical interactions: \emph{real-time} classical computations that must be performed within the coherence times of the qubits, and \emph{near-time} computations with less stringent timing. The real-time computations are critical for error correction and circuits that take advantage of feedback or feedforward. However, the demands of the real-time, low-latency domain might impose considerable restrictions on the available compute resources in that environment. For example real-time computations might be constrained by limited memory or reduced clock speeds. Whereas in the near-time domain we assume more generic compute resources, including access to a broad set of libraries and runtimes. Since the near-time domain is already adequately described by existing programming tools and frameworks, we choose in OpenQASM\,3 to focus on the real-time domain which must be more tightly coupled to the execution of gates and measurement on qubits.

We use a dataflow model \cite{dataflow04} where each instruction may be executed by an independent
processing unit when its input data becomes available. Concurrency and parallelism are essential
features of quantum control hardware. For example, parallelism is necessary for a fault-tolerance
accuracy threshold to exist \cite{steane99,abo08}, and we expect concurrent quantum and classical
processing to be necessary for quantum error-correction~\cite{fwh12,delfosse20,das20}.

Our extended OpenQASM language expresses a quantum circuit, or any equivalent representation of it, as a collection of (quantum) basic blocks and flow control instructions. This differs from the role of a high-level language. High-level languages may include mechanisms for quantum memory management and garbage collection, quantum datatypes with non-trivial semantics, or features to specify classical reversible programs to synthesize into quantum oracle implementations. Optimization passes at this high level work with families of quantum operations whose specific parameters might not be known yet. These passes could be applied once at this level and benefit every program instance we execute later. High-level intermediate representations may differ from OpenQASM until the point in the workflow where a specific circuit is generated as an instance from the family.

On the other hand, while quantum assembly languages are considered low-level programming languages, they are not often directly consumed by quantum control hardware. Rather, they sit at multiple levels in the middle of the stack where they might be hand-written, generated by scripts (meta-programming), or targeted by higher-level software tools and then are further compiled to lower-level instructions that operate at the level of analog signals for controlling quantum systems.

Our choice of features to add to OpenQASM is guided by use cases. Although OpenQASM is not a high-level language, many users would like to write simple quantum circuits by hand using an expressive domain-specific language. Researchers who study circuit compiling need high-level information
recorded in the intermediate representations to inform the optimization and synthesis algorithms.
Experimentalists prefer the convenience of writing circuits at a relatively high level but often
need to manually modify timing or pulse-level gate descriptions at various points in the circuit. Hardware
engineers who design the classical controllers and waveform generators prefer languages that are
practical to compile given the hardware constraints and make explicit circuit structure that the controllers can take advantage of. Our choice of language features is intended to acknowledge all of these potential audiences.

What follows is a review of the major new features introduced in OpenQASM\,3 with a focus on explaining the ideas that motivated those changes and the design intent of how the features work. This manuscript is not a formal specification of the language. Rather, the formal specification exists as a live document~\cite{livedoc} backed by a GitHub repository where people may comment or suggest changes through issues and pull requests. While this document hopefully serves as a useful guide to understanding OpenQASM\,3, readers interested in writing circuits with compliant syntax are strongly encouraged to refer to the specification~\cite{livedoc}. To facilitate continued evolution and refinement of the language, a governance structure has been established~\cite{openqasmgovernance} including a technical steering committee to formalize how changes will be accepted and to spur the development of working groups to study specific issues of language design.

\section{Design philosophy and execution model}
 
In OpenQASM\,3, we aim to describe a broader set of quantum circuits with concepts beyond simple qubits and gates.
Chief among them are arbitrary classical control flow, gate modifiers (e.g. control and inverse), timing, and microcoded pulse implementations. While these extensions are not strictly necessary from a theoretical point of view --- any quantum computation could in principle be described using OpenQASM\,2 --- in practice they greatly expand the expressivity of the language.

\subsection{Versatility in describing logical circuits}

One key motivation for the new language features is the ability to describe new kinds of circuits and experiments. For example, repeat-until-success algorithms~\cite{rus} or magic state distillation protocols~\cite{msd} have non-deterministic components that can be programmed using the new classical control flow instructions. There are other examples which show benefits to circuit width or depth from incorporating classical operations. For instance, with measurement and feedforward any Clifford operation can be executed in constant depth~\cite{josza2005}. The same is true of the quantum Fourier transform~\cite{whitmeyer}. This extension also allows description of teleportation and feedforward in the measurement model~\cite{raussendorf2001, josza2005}. This collection of examples suggests the potential of the extended quantum circuit family to reduce the requirements needed to achieve quantum advantage in the near term.

It is not our intent, however, to transform OpenQASM into a general-purpose programming language suitable for classical programming. A general quantum application includes classical computations that need to interact with quantum hardware. For instance, some applications require pre-processing of problem data, such as choosing a co-prime in Shor's algorithm~\cite{shor1994}, post-processing to compute expectation values of Pauli operators, or further generation of circuits like in the outer loop of many variational algorithms~\cite{vqe}. However, these classical computations do not have tight deadlines requiring execution within the coherence time of the quantum hardware. Consequently, we think of their execution in a \emph{near-time} context which is more conveniently expressed in existing classical programming frameworks. Our classical computation extensions in OpenQASM\,3 instead focus on the \emph{real-time} domain which must be tightly coupled to quantum hardware. Even in the real-time context we enable re-use of existing tooling by allowing references to externally specified real-time functions defined outside of OpenQASM itself.

\subsection{Versatility in the level of description for circuits}

We wish to use the same tools for circuit development \emph{and} for lower-level control sequences needed for calibration, characterization, and error mitigation. Hence, we need the ability to control timing and to connect quantum instructions with their pulse-level implementations for various qubit modalities. For instance, dynamical decoupling \cite{viola1999} or characterizations of decoherence and crosstalk \cite{gambetta2012} are all sensitive to timing, and can be programmed using the new timing features. Pulse-level calibration of gates can be fully described in OpenQASM\,3.

Another design goal for the language is to create a suitable intermediate representation (IR) for quantum circuit compilation. Recognizing that quantum circuits have multiple levels of specificity from higher-level theoretical computation to lower-level physical implementation, we envision OpenQASM as a multi-level IR where different levels of abstraction can be used to describe a circuit. We can speak broadly of two levels of abstraction: logical and physical levels. A logical-level circuit (see Section~\ref{sec:logical-features}) is described by abstract, discrete-time gates and real-time classical computations. This is lowered by a compiler to a physical-level circuit (see Section~\ref{sec:physical-features}) where quantum operations are implemented by continuous, time-varying signals with specific timing constraints. We introduce semantics that allow the compiler to optimize and re-write the circuit at different levels. For example, the compiler can reason directly on the level of controlled gates or gates raised to some power without decomposing these operations, which may obscure some opportunities for optimization. Furthermore, the new semantics allow capturing intent in a portable manner without tying it to a particular implementation. For example, the timing semantics in OpenQASM are designed to allow higher-level descriptions of timing constraints without being specific about individual gate durations.

\subsection{Continuity with OpenQASM\,2}

In the design of interfaces for quantum computing, particularly at the early stage that we are currently in, backwards compatibility is not to be prized above all else.
However, in recognition of the current importance of the preceding version of OpenQASM as a \emph{de facto} standard, we made design choices to introduce as few  incompatibilities between OpenQASM\,2 and OpenQASM\,3 as practically possible.

Our intent is that an OpenQASM\,2 circuit can be interpreted as an OpenQASM\,3 circuit with no change in meaning.
In doing so, we hope to introduce as little disruption to software projects which rely on OpenQASM, and to make OpenQASM\,3 easier to adopt for those software developers who are already familiar with OpenQASM\,2.
We describe the supported features of OpenQASM\,2 in more detail in Section~\ref{sec:backwards-compatibility}.

\begin{figure*}
\centering
\includegraphics[width=.9\textwidth]{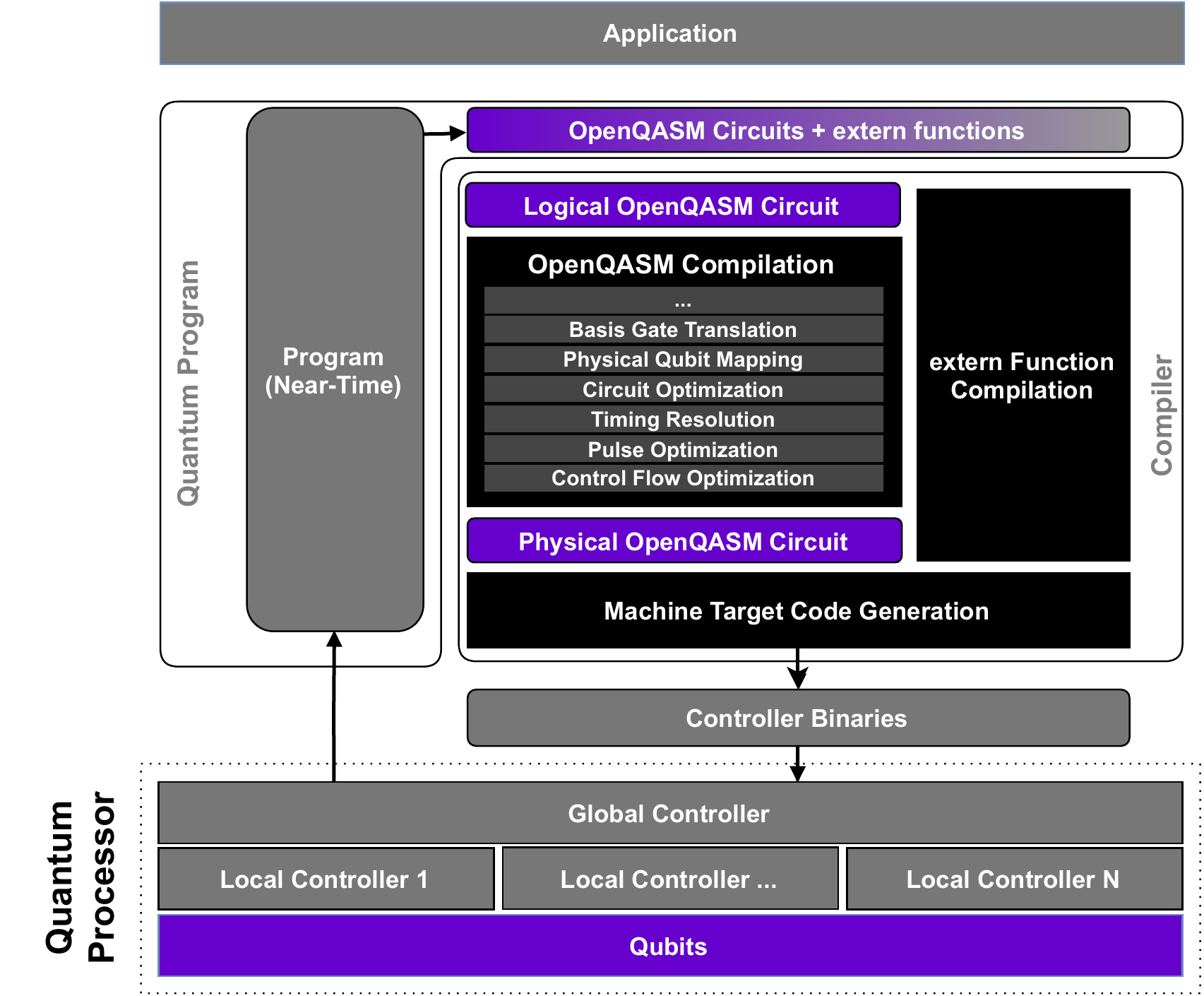}
\caption{The compilation and execution model of a quantum program, and OpenQASM's place in the flow. An application is written in terms of a quantum program. Certain regions of a quantum program are purely classical and can be written in classical programming languages for execution in a near-time context. The quantum program emits payloads for execution on a quantum processor (QPU). This payload consists of extended quantum circuits and external real-time classical functions. OpenQASM is the language to describe the quantum circuits, which includes interface calls to the external classical functions. There may be higher-order aspects of the circuit payload which are optimized before OpenQASM is generated. An OpenQASM compiler can transform and optimize all aspects of the circuits described with the IR, including basis gates used, qubit mapping, timing, pulses, and control flow. The final physical circuit plus extern functions are passed to a target code generator which produces binaries for the QPU.}
\label{fig:program-flow}
\end{figure*}

\subsection{Execution model}

We illustrate a potential application compilation and execution flow for quantum programs in Figure~\ref{fig:program-flow}. An application might be composed of several quantum programs including near-time calculations and quantum circuits. The quantum program interacts with quantum hardware by emitting OpenQASM3 circuits and external real-time functions. Initial generation of these circuits and \qasmline{extern} functions requires a deep compilation stack to transform the circuits and \qasmline{extern}s into a form that is executable on the quantum processor (QPU). We represent these multiple stages by logical OpenQASM and physical OpenQASM. The logical OpenQASM represents the intent of the extended quantum circuit whereas the physical OpenQASM is the lowest level where the circuit is mapped and scheduled on specific qubits. A quantum program might work with even higher-level circuit representations than those expressible with OpenQASM\,3, but in this flow, manipulations of these higher-order objects would be contained within the quantum program. A quantum program is not constrained to produce only logical OpenQASM. Sometimes it might emit physical OpenQASM, or later stages of a program might take advantage of lower-cost ways to produce binaries executable by the QPU. For instance, a program describing the optimization loop of a variational algorithm might directly manipulate a data section in the binaries to describe circuits with updated circuit parameters~\cite{Karalekas_2020}. We generically expect quantum programs to be able to enter the compilation toolchain at varying levels of abstraction as appropriate for different phases of program execution.

Our language model further assumes that the QPU has a global controller that orchestrates execution of the circuit. This global controller centralizes control flow decisions and has the ability to execute \qasmline{extern} computations emitted by the quantum program. Some QPUs may also have a collection of local controllers that interact with a subset of qubits. Such segmentation of the QPU could enable concurrent execution of independent code segments.

\section{Comparison to OpenQASM\,2}
\label{sec:backwards-compatibility}

OpenQASM\,3 is designed to allow quantum circuits to be expressed in greater detail, and also in a more versatile way.
However, we considered it important in extending this functionality to remain as close to the language structure of OpenQASM\,2 as practically possible.

Throughout this article, we occasionally note similarities or differences between OpenQASM versions 2 and~3.
In this section, for the convenience of developers who are familiar with OpenQASM\,2, we explicitly describe the OpenQASM\,2 features which developers may rely on to work identically in OpenQASM\,3.

\vspace*{-.5ex}
\subsubsection*{Circuit header}
Top-level circuits in OpenQASM\,2 must begin with the statement ``\qasmline{OpenQASM 2.x;}'',%
    \footnote{%
        Throughout the rest of the article, to reduce clutter, we often omit terminal semicolons for in-line examples.
    }
possibly preceded by one or more blank lines or lines of comments, where \qasmline{x} is a minor version number (such as \qasmline{0}).
Other source-files could be included using the syntax \qasmline{include "filename"}, in the header or elsewhere in the circuit.
Top-level circuits in OpenQASM\,3 begin similarly, \emph{e.g.},~with a statement ``\qasmline{OpenQASM 3.0}''; and OpenQASM\,3 supports \qasmline{include} statements in the same way.

\vspace*{-.5ex}
\subsubsection*{Circuit execution}

The top-level of an OpenQASM\,2 circuit is given by a sequence of instructions at the top-level scope.
No means of specifying an explicit entry point is provided: the circuit is interpreted as a stream of instructions, with execution starting from the first instruction after the header.
A top-level OpenQASM\,3 circuit may also be provided in this way: as a sequence of instructions at the top-level scope, without requiring an explicit entry point.

\vspace*{-.5ex}
\subsubsection*{Quantum and classical registers}
In OpenQASM\,2, the only storage types available are qubits which are allocated as part of a \qasmline{qreg} declaration, and classical bits which are allocated as part of a \qasmline{creg} declaration, for example:
\begin{qasmblock}
qreg q[5]; // a register 'q' with five qubits
creg c[2]; // a register 'c' with two classical bits
\end{qasmblock}
This syntax is supported in OpenQASM\,3, and is equivalent to the (preferred%
    \footnote{%
        The \qasmline{qreg} and \qasmline{creg} keywords may not be supported in future versions of OpenQASM.
    })
syntax
\begin{qasmblock}
qubit[5] q; // a register 'q' of five qubits
bit[2] c;   // a register 'c' of two classical bits
\end{qasmblock}
which respectively declares \qasmline{q} and \qasmline{c} as registers of more primitive \qasmline{qubit} or \qasmline{bit} storage types.
(OpenQASM\,3 also introduces other storage types for classical data: see Section~\ref{sec:rt-classical}.)

\vspace*{-.5ex}
\subsubsection*{Names of gates, variables, and constants}
In OpenQASM\,2, the names of registers and gates must begin with a lower-case alphabetic ASCII character.
This constraint is relaxed in OpenQASM\,3: identifiers may now begin with other characters, such as capital letters, underscores, and a range of  unicode characters.
For example, angular values used as gate arguments may now be represented by greek letters.
(The identifier \qasmline{π} is reserved in OpenQASM\,3 to represent the same constant as \qasmline{pi}.)
Note that OpenQASM\,3 has a slightly different set of keywords from OpenQASM\,2.
This may cause errors in OpenQASM circuits which happen to use a keyword from version~3 as an identifier name: such identifiers would have to be renamed in order for the source to be valid OpenQASM\,3.

\vspace*{-.5ex}
\subsubsection*{Basic operations}
In OpenQASM\,2, there were four basic instructions which affected stored quantum data:
\begin{itemize}
\item
    Single-qubit unitaries specified with the syntax  \qasmline{U(a,b,c)} for some angular parameters \qasmline{a}, \qasmline{b}, and \qasmline{c}, acting on a single-qubit argument.
    This syntax is also supported in OpenQASM\,3, and specifies an equivalent unitary transformation.%
        \footnote{%
            The OpenQASM\,3 specification for single-qubit unitaries, described in Eq.~\ref{eq:defineU}, differs by a global phase from the specification in OpenQASM\,2.
            This change has no effect on the results of an OpenQASM\,2 circuits.
        }

\item
    Two-qubit controlled-NOT operations, using the keyword \qasmline{CX}, acting on two single-qubit arguments.
    In OpenQASM\,3, the \qasmline{CX} instruction is \emph{no longer} a basic instruction or a keyword, but may be defined from more primitive commands.
    To adapt an OpenQASM\,2 circuits to be valid OpenQASM\,3, one may add a single instruction to define \qasmline{CX} using a ``\qasmline{ctrl @}'' gate modifier:
    \begin{qasmblock}
    gate CX c, t { ctrl @ U(π,0,π) c, t }
    \end{qasmblock}
    A~definition of this sort for \qasmline{CX} is also provided in the standard gate library for OpenQASM\,3, which we describe in Section~\ref{sec:std-gate-library}.
    (The meaning of the ``\qasmline{ctrl @}'' gate modifier is described in Section~\ref{sec:ctrl-modifier}.)
\item
    Non-unitary \qasmline{reset} and \qasmline{measure}  operations, acting on a single-qubit argument (and producing a single bit outcome in the case of \qasmline{measure}).
    The statement \qasmline{reset q[j]} operation has the effect of discarding the data stored in \qasmline{q[j]}, replacing it with the state $\ket{0}$.
    The statement \qasmline{measure q[j] -> c[k]} has the effect of measuring \qasmline{q[j]} in the standard basis, \emph{i.e.},~projecting the state into one of the eigenstates of the Pauli $Z$ operator, and storing the corresponding bit-value in a classical bit \qasmline{c[k]}.
    OpenQASM\,3 supports these operations without changes, and also supports the alternative syntax \qasmline{c[k] = measure q[j]} for measurements.
\end{itemize}

\vspace*{-.5ex}
\subsubsection*{Gate declarations}
OpenQASM\,2 supports \qasmline{gate} declarations to specify user-defined gates.
The definitions set out a fixed number of single-qubit arguments, and optionally some arguments which are taken to be angular parameters.
These declarations use syntax such as%
\begin{qasmblock}
gate h a { U(π/2, 0, π) a; }
gate p(θ) a { U(0, 0, θ) a; }
gate cp(θ) a, b { p(θ/2) a; p(θ/2) b; CX a,b; p(-θ/2) b; CX a,b; }
\end{qasmblock}
The specification of these gates, in the code-blocks enclosed by braces \texttt{\{ $\ldots$ \}}, are by sequences of \emph{unitary} operations, whether basic unitary operations or ones defined by earlier \qasmline{gate} declarations.
OpenQASM\,3 supports \qasmline{gate} declarations with this syntax, and also admits only unitary operations as part of the gate definitions --- but has greater versatility in how those unitary operations may be described (see Section~\ref{sec:modifiers}), and also provides other ways to define subroutines (see Section~\ref{sec:rt-classical}).

\vspace*{-.5ex}
\subsubsection*{Implicit iteration}
Operations on one or more qubits could be repeated across entire registers as well, using implicit iteration.
For instance, for any operation \mbox{``\qasmline{qop q[j]}''} which acts on one qubit \qasmline{q[j]}, we can instead perform it independently on every qubit in \qasmline{q} by omitting the index (as in ``\qasmline{qop q}'').
For operations on two qubits of the form \mbox{``\qasmline{qop q[j], r[k]}''}, one or both arguments could omit the index, in which case the instruction would be repeated either for $\texttt{j}\!=\!1,2,\ldots$ or $\texttt{k}\!=\!1,2,\ldots$ (if one index is omitted), or for $\texttt{j}\!=\!\texttt{k}\!=\!1,2,\ldots$ (if both indices are omitted).
This functionality also extends to operations on three qubits or more.
OpenQASM\,3 supports this functionality unchanged, for \qasmline{gate} subroutines.
(OpenQASM\,3 also supports other kinds of subroutine beyond \qasmline{gate} definitions, and these other subroutines do not support implicit iteration in this way.)

\vspace*{-.5ex}
\subsubsection*{Control flow}
The only control flow supported by OpenQASM\,2 are \qasmline{if} statements.
These can be used to compare the value of a classical bit-register (interpreted as a little-endian representation of an integer) to an integer, and conditionally execute a single gate.
An example of such a statement is
\begin{qasmblock}
if (c == 5) mygate q, r, s;
\end{qasmblock}
which would test whether a classical register \qasmline{c} (of length three or more) stores a bit-string $c_{n{-}1} \cdots c_3  c_2  c_1 c_0$ equal to $\texttt{0} \cdots \texttt{0101}$, to determine whether to execute \qasmline{mygate q, r, s}.
If-statements (and classical register comparisons) of this kind are supported in OpenQASM\,3, as is more general syntax for \qasmline{if} statements and other forms of control-flow (see Section~\ref{sec:rt-classical}).

\vspace*{-.5ex}
\subsubsection*{Barrier instructions}
OpenQASM\,2 provides a \qasmline{barrier} instruction which may be invoked with or without arguments.
When invoked with arguments, either of individual qubits or whole quantum registers, it instructs the compiler not to perform any optimizations that involve moving or simplifying operations acting on those arguments, across the source line of the \qasmline{barrier} statement;
when invoked without arguments, it has the same effect as applying it to all of the quantum registers that have been defined.
This operation is also supported in OpenQASM\,3, with the same meaning.

\vspace*{-.5ex}
\subsubsection*{Opaque definitions}
OpenQASM\,2 supports \qasmline{opaque} declarations, to declare gates whose physical implementation may be possible but is not readily expressed in terms of unitary gates.
OpenQASM\,3 provides means of defining operations on a lower level than unitary gates (see Section~\ref{sec:calibrating}), therefore opaque definitions are not needed.
OpenQASM\,3 compilers will simply ignore any \qasmline{opaque} declarations.

\vspace*{-.5ex}
\subsubsection*{Circuit output}

The outputs of an OpenQASM\,2 circuit are the values stored in any declared classical registers.
OpenQASM\,3 introduces a means of explicitly declaring which variables are to be produced as output or taken as input (see Section~\ref{sec:parameterised-circuits}).
However, any OpenQASM\,3 circuit which does not specify either an output or input will, by default, also produce all of its classical stored variables (whether of type \qasmline{creg}, or a different type) as output.

\section{Concepts of the language: the logical level}
\label{sec:logical-features}

OpenQASM\,3 is a multi-level IR for quantum computations, which expresses concepts at both a logical and a more fine-grained physical level.
In this section we give a overview of the important features of the logical level of OpenQASM\,3, presenting the more low-level features in Section~\ref{sec:physical-features}. For a finer-grained specification, we direct readers to the live specification \cite{livedoc}.

\subsection{Continuous gates and hierarchical library}
\label{sec:gates}

We define a mechanism for parameterizing unitary matrices to define quantum gates.  The parameterization uses a set of built-in single-qubit gates and a gate modifier to construct a controlled version to generate a universal gate set \cite{barenco95}. Early QASM languages assumed a discrete set of quantum gates, but OpenQASM is flexible enough to describe universal computation with a continuous gate set. This gate set was chosen for the convenience of defining new quantum gates and is not an enforced compilation target. Instead of allowing the user to write a unitary as an $n \times n$ matrix, we make gate definitions through hierarchical composition allowing for code reuse to define more complex operations \cite{scaffold,dousti16}. For many gates of practical interest, there is a circuit representation with a polynomial number of one- and two-qubit gates, giving a more compact representation than requiring the programmer to express the full $n \times n$ matrix. In the worst case, a general $n$-qubit gate can be defined using an exponential number of these gates.

We now describe this built-in gate set. Single-qubit unitary gates are parameterized as
\begin{equation}\label{eq:defineU}
U(\theta,\phi,\lambda) := \left(\begin{array}{cc}
    \cos(\theta/2) & -e^{i\lambda}\sin(\theta/2) \\
e^{i\phi}\sin(\theta/2) & e^{i(\phi+\lambda)}\cos(\theta/2) \end{array}\right)
\end{equation}
This expression specifies any element of $U(2)$ up to a global phase. The global phase is here chosen so that the upper left matrix element is real. For example, \qasmline{U(π/2, 0, π) q[0];}
applies the Hadamard gate
$H = \frac{1}{\sqrt{2}} \begin{psmallmatrix}
    1 & \phantom-1 \\
    1 & -1\end{psmallmatrix}$
to qubit \qasmline{q[0]}.
For the convenience of the reader, we note two important decompositions of $U$, through an Euler angle decomposition using rotations $R_z(\alpha) = \exp(-i\alpha Z/2)$ and $R_y(\beta) = \exp(-i\beta Y/2)$, and a $\{H, P\}$ basis representation where $P(\alpha) =  \begin{psmallmatrix}
    1 \,&\, 0 \\
    0 \,&\, e^{i\alpha\!}\end{psmallmatrix}$:
\begin{equation}\label{eq:decompU}
U(\theta,\phi,\lambda) \;=\; e^{i(\phi+\lambda)/2}\, R_z(\phi)R_y(\theta)R_z(\lambda) \;=\; e^{i\theta/2} \, P(\phi \!+\!\pi/2) H P(\theta) H P(\lambda\!-\!\pi/2).
\end{equation}

\medskip
\noindent\textbf{Note.~}
Users should be aware that the definition in Eq.~\ref{eq:defineU} is scaled by a factor of $e^{i(\phi+\lambda)/2}$ when compared to the original definition in OpenQASM\,2~\cite{CBSG17}. This implies that the transformation described by \,\qasmline{U(a,b,c)}\, has determinant $e^{i(\texttt{a}+\texttt{c})}$, and is not in general an element of $SU(2)$. This change in global phase has no observable consequence for OpenQASM\,2 circuits. We also note that \qasmline{U} is the only uppercase keyword in OpenQASM\,3.
\smallskip

New gates are associated to a unitary transformation by defining them using a sequence of built-in or previously defined gates and gate modifiers. For example, the \qasmline{gate} declaration
\begin{qasmblock}
gate h q { U(π/2, 0, π) q; }
\end{qasmblock}
defines a new gate called ``\qasmline{h}'' and associates it to the unitary matrix of the Hadamard gate. Once we have defined ``\qasmline{h}'', we can use it in later \qasmline{gate} blocks. The definition does not necessarily imply that \qasmline{h} is implemented by an instruction \qasmline{U(π/2, 0, π)} on the quantum computer. The implementation is left to the user and/or compiler (see Section~\ref{sec:calibrating}), given information about the instructions supported by a particular target.

Controlled gates can be constructed by attaching a control modifier to an existing gate. For example, the NOT gate (\emph{i.e.},~the Pauli $X$ operator) is given by \qasmline{U(π, 0, π)} and the block
\begin{qasmblock}
gate CX c, t { ctrl @ U(π, 0, π) c, t; }
CX q[1], q[0];
\end{qasmblock}
defines the gate
\begin{equation}
\texttt{CX} := I\oplus \texttt{x} = \left(\begin{array}{cccc}
1 & 0 & 0 & 0 \\
0 & 1 & 0 & 0 \\
0 & 0 & 0 & 1 \\
0 & 0 & 1 & 0 \end{array}\right)
\end{equation}
and applies it to \qasmline{q[1]} and \qasmline{q[0]}. This gate applies a bit-flip to \qasmline{q[0]} if \qasmline{q[1]} is one, leaves \qasmline{q[1]} unchanged if \qasmline{q[0]} is zero, and acts coherently over superpositions. The control modifier is described in more detail in Section~\ref{sec:modifiers}.

\medskip
\noindent
\textbf{Remark.~} Throughout the document we use a tensor order with higher index qubits on the left. This ordering labels states in a way that is consistent with how numbers are usually written with the most significant digit on the left. In this tensor order, \qasmline{CX q[0], q[1];} is represented by the matrix
\begin{equation}
\left(\begin{array}{cccc}
1 & 0 & 0 & 0 \\
0 & 0 & 0 & 1 \\
0 & 0 & 1 & 0 \\
0 & 1 & 0 & 0 \end{array}\right).
\end{equation}
\medskip

From a physical perspective, any unitary $U$ is indistinguishable from another unitary $\mathrm e^{i\gamma} U$ which only differs by a global phase. When we attach a control to these gates, however, the global phase becomes a relative phase that is applied when the control qubit is one. To capture the programmer's intent, a built-in global phase gate allows the inclusion of arbitrary global phases on circuits. The instruction \qasmline{gphase(γ)} adds a global phase of $\mathrm e^{i\gamma}$ to the scope containing the instruction. For example,
\begin{qasmblock}
gate rz(τ) q { gphase(-τ/2); U(0, 0, τ) q; }
ctrl @ rz(π/2) q[1], q[0];
\end{qasmblock}
constructs the gate
\begin{equation}
    R_z(\tau)
    =
    \exp(-i\tau Z/2)
    =
    \begin{pmatrix}
        \mathrm e^{-i\tau/2} & 0 \\
        0 & \mathrm e^{i\tau/2}
    \end{pmatrix}
    =
    \mathrm e^{-i\tau/2}
    \begin{pmatrix}
        1 & 0 \\
        0 & \mathrm e^{i\tau}
    \end{pmatrix}
\end{equation}
and applies the controlled form of that phase rotation gate
\begin{equation}
\setlength{\arraycolsep}{2ex}
    I\oplus R_z(\pi/2)
=
    \begin{pmatrix}
        1 & 0 & 0 & 0 \\
        0 & 1 & 0 & 0 \\
        0 & 0 & \!\mathrm e^{-i\pi/4}\! & 0 \\
        0 & 0 & 0 & \!\mathrm e^{i\pi/4}\!
    \end{pmatrix}.
\end{equation}

In OpenQASM\,2, \qasmline{gate} subroutines are defined in terms of built-in gates \qasmline{U} and \qasmline{CX}. In OpenQASM\,3, we can define controlled gates using the control modifier, so it is no longer necessary to include a built-in \qasmline{CX} gate. For backwards compatibility, we include the gate \qasmline{CX} in the standard gate library (described below), but it is no longer a keyword of the language.

\subsubsection{Standard gate library}
\label{sec:std-gate-library}

We define a standard library of OpenQASM\,3 gates in a file we call \texttt{stdgates.inc}. Any OpenQASM\,3 circuit that includes the standard library can make use of these gates.

\qasmfile{examples/stdgates.inc}

\subsection{Gate modifiers}\label{sec:modifiers}

We introduce a mechanism for modifying existing unitary gates \qasmline{g} to define new ones. A modifier defines a new unitary gate from \qasmline{g}, acting on a space of equal or greater dimension, that can be applied as an instruction. We add modifiers for inverting, exponentiating, and controlling gates. This can be useful for programming convenience and readability, but more importantly it allows the language to capture gate semantics at a higher level which aids in compilation. For example, optimization opportunities that exist by analysing controlled unitaries may be exceedingly hard to discover once the controls are decomposed.

\subsubsection*{The control modifier}
\label{sec:ctrl-modifier}

The modifier \qasmline{ctrl @ g} represents a controlled-\qasmline{g} gate with one control qubit. This gate is defined on the tensor product of the control space and target space by the matrix $I\oplus \texttt{\qasmline{g}}$\,, where $I$ has the same dimensions as \qasmline{g}.
(We regard \qasmline{ctrl @ gphase(a)} as a special case, which is defined to be the gate $1 \oplus e^{i\;\!\mathtt{a}} = \text{\qasmline{U(0,0,a)}}$.)
A compilation target may or may not be able to execute the controlled gate without further compilation. The user or compiler may rewrite the controlled gate to use additional ancillary qubits to reduce the size and depth of the resulting circuit.

For example, we can define the Fredkin (or controlled-swap) gate in two different ways:
\begin{qasmblock}
// Some useful reversible gates
gate x q { U(π, 0, π) q; }
gate cx c, t { ctrl @ x c, t; }
gate toffoli c0, c1, t { ctrl @ cx c0, c1, t; }
gate swap a, b { cx a, b; cx b, a; cx a, b; }

// Fredkin definition #1
gate fredkin1 c, a, b {
    cx b, a;
    toffoli c, a, b;
    cx b, a;
}

// Fredkin definition #2
gate fredkin2 c, a, b { ctrl @ swap c, a, b; }
\end{qasmblock}
The definitions are equivalent because they result in the same unitary matrix for both Fredkin gates. If we directly apply the definitions, \qasmline{fredkin1} involves fewer controlled operations than \qasmline{fredkin2}. A compiler pass may infer this fact or even rewrite \qasmline{fredkin2} into \qasmline{fredkin1}. Note that although the \qasmline{toffoli} gate is well defined here, the compiler or user needs to do further work to synthesize it in terms of more primitive operations, such as 1- and 2-qubit gates, or pulses (see Section~\ref{sec:calibrating}).

For a second example, if we go one step further, we can define a controlled Fredkin gate.
If we use the $ABA^\dagger$ pattern in the definition of the \qasmline{fredkin1} gate, we require only one controlled \qasmline{toffoli} gate after we remove unneeded controls from the \qasmline{cx} gates. Although the controlled \qasmline{toffoli} gate is defined on four qubits, it may be beneficial for a compiler or user to synthesize the gate using extra qubits as scratch space to reduce the total gate count or circuit depth, as in the following example:
\begin{qasmblock}
// Require: the scratch qubit is 'clean', i.e. initialized to zero
// Ensure: the scratch qubit is returned 'clean'
gate cfredkin2 c0, c1, a, b, scratch {
    cx b, a;                  // (from fredkin1)
    toffoli c0, c1, scratch;  // implement ctrl @ toffoli c0, c1, a, b
    toffoli scratch, a, b;    // by computing (and uncomputing) an AND of the two controls
    toffoli c0, c1, scratch;  // in the scratch qubit
    cx b, a;                  // (from fredkin1)
}
\end{qasmblock}

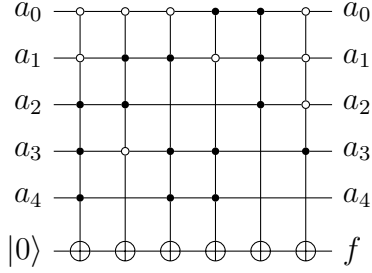
\begin{figure*}
    \centering
    \begin{tikzpicture} 
        \begin{yquant} 
        qubit {$a_\idx$} a[5];
        qubit {$\ket{0}$} f;
        not f | a[2-4] ~a[0, 1]; 
        not f | a[1-2] ~a[0],a[3]; 
        not f | a[1],a[3-4] ~a[0]; 
        not f | a[0],a[3-4] ~a[1]; 
        not f | a[0-2]; 
        not f | a[3], ~a[0-2]; 
        output {$a_\idx$} a;
        output {$f$} f;
        \end{yquant} 
    \end{tikzpicture}
    \caption{\label{fig:boolean-function}
    Toffoli network to compute a Boolean function
    $f = \bar{a}_0\bar{a}_1a_2a_3a_4 \oplus
         \bar{a}_0a_1a_2\bar{a}_3 \oplus
         \bar{a}_0a_1a_3a_4 \oplus
         a_0\bar{a}_1a_3a_4 \oplus
         a_0a_1a_2 \oplus
         \bar{a}_0\bar{a}_1\bar{a}_2a_3$
    in a target qubit.
    The control pattern may be expressed in OpenQASM via control gate modifiers.}
\end{figure*}

We further provide the \qasmline{negctrl} modifier to control operations with \emph{negative} polarity, \emph{i.e.},~conditioned on the control bit being zero rather than one. Both \qasmline{ctrl} and \qasmline{negctrl} optionally accept an argument $n$ which specifies the number of controls of the appropriate type, which are added to the front of the list of arguments (omission means $n=1$). In each case, $n$ must be a positive integer expression which is a `compile-time constant'.%
        \footnote{%
            By `compile-time constant', we mean an expression which is actually constant, or which depends only on \textbf{(a)}~variables which take fixed values, or \textbf{(b)}~iterator variables in \qasmline{for} loops which can be unrolled by the compiler (\emph{i.e.,}~which have initial and final iterator values which are themselves compile-time constants).
        }
For example, this can be useful in writing classical Boolean functions~\cite{schmitt2019scaling}, as demonstrated by the following example (illustrated in Figure~\ref{fig:boolean-function}).

\qasmfile{examples/boolean.qasm}

\subsubsection*{The inversion modifier}

The modifier \qasmline{inv @ g} represents the inverse \qasmline{g}$^\dagger$ of the gate \qasmline{g}.
This can be easily computed from the definition of the gate \qasmline{g} as follows.
\begin{itemize}
\item
    The inverse of any unitary operation $U=U_mU_{m-1}\dots U_1$ can be defined recursively by reversing the order of the gates in its definition and replacing each of those with their inverse $U^\dagger=U_1^\dagger U_2^\dagger\dots U_m^\dagger$.
\item
    The inverse of a controlled operation $I\oplus U$ can be defined by reversing the operation which is controlled, as  $(I\oplus U)^\dagger=I\oplus U^\dagger$ for any unitary $U$.
    That is, \mbox{\qasmline{inv @ ctrl @ g}} is defined to have the same meaning as \mbox{\qasmline{ctrl @ inv @ g}} (commuting the \mbox{`\qasmline{inv @}'} modifier past the \mbox{`\qasmline{ctrl @}'} modifier).
\item
    The base case is given by replacing \qasmline{inv @ U(a, b, c)} by \qasmline{U(-a, -c, -b)}, and replacing \qasmline{inv @ gphase(a)} by \qasmline{gphase(-a)}.
\end{itemize}
For example,
\begin{qasmblock}
gate rz(τ) q { gphase(-τ/2); U(0, 0, τ) q; }
inv @ rz(π/2) q[0];
\end{qasmblock}
applies the gate \qasmline{inv @ rz(π/2)}, which is defined by \qasmline{{ gphase(π/4); U(0,-π/2,0); }}\,.
For an additional example: the gate \mbox{\,\qasmline{inv @ ctrl @ rz(π/2) q[1], q[0]}\,} would be rewritten as \mbox{\,\qasmline{ctrl @ inv @ rz(π/2) q[1], q[0]}\,}, which is the same as \mbox{\qasmline{ctrl @  rz(-π/2) q[1], q[0]}\,}.

\subsubsection*{The powering modifier}

The modifier \qasmline{pow(r) @ g} represents the $r$th power \qasmline{g}$^r$ of the gate \qasmline{g}, where $r$ is an integer or floating point number. Every unitary matrix $U$ has a unique principal logarithm $\log U = iH$, where $H$ has only real eigenvalues $E$ which satisfy $-\pi< E \leq \pi$. For any real $r$, we define the $r$th power as $U^r = \exp(i r H)$.

For example, consider the Pauli $Z$ operator $\textrm{\qasmline{z}} = \begin{psmallmatrix}
1 & 0 \\ 0 & -1\end{psmallmatrix}$.
The principal logarithm is $\log \textrm{\qasmline{z}} =\begin{psmallmatrix}
0 & 0 \\ 0 & i\pi \end{psmallmatrix}$, from which it follows that \qasmline{z}$^{1/2} = \exp(\tfrac{1}{2} \log \texttt{\qasmline{z}}) = \begin{psmallmatrix} 1 & 0 \\ 0 & i \end{psmallmatrix}$.
In this way, the definitions
\begin{qasmblock}
gate z q { U(0, 0, π) q; }
gate s q { pow(1/2) @ z q; }
\end{qasmblock}
allow us to define the gate \qasmline{s} as the square root of \qasmline{z}.
(A compiler pass is responsible for resolving $S$ into gates.)
For another example, the principal logarithm of the Pauli $X$ operator $\textrm{\qasmline{x}} = \begin{psmallmatrix}
0 & 1 \\ 1 & 0\end{psmallmatrix}$
is $\log \textrm{\qasmline{x}}=\tfrac{i\pi}{2}\begin{psmallmatrix} +1 \,&\, -1 \\ -1 \,&\, +1\end{psmallmatrix}$.
One can confirm that this operator has eigenvalues $0$ and $i\pi$, and $\textrm{\qasmline{x}}^{1/2} := \exp(\tfrac{1}{2}\log \texttt{\qasmline{x}}) = \tfrac{1}{2}\begin{psmallmatrix} 1{+}i \,&\, 1{-}i \\ 1{-}i \,&\, 1{+}i \end{psmallmatrix}$. Therefore the definitions of the $\sqrt{X}$ gate
\begin{qasmblock}
gate sx a { gphase(π/4); U(π/2, -π/2, π/2) a; }
gate sx a { pow(1/2) @ x a; }
\end{qasmblock}
are equivalent.

In the case where $r$ is an integer, the $r$th power \qasmline{pow(r) @ g} can be implemented simply (albeit less efficiently) as $r$ repetitions of \qasmline{g} when $r>0$, or $r$ repetitions of \qasmline{inv @ u} when $r<0$. Continuing our previous example,
\begin{qasmblock}
pow(-2) @ s q[0];
\end{qasmblock}
defines and applies \qasmline{s}$^{-2}$. This can be compiled to
\begin{qasmblock}
inv @ pow(2) @ s q[0];
\end{qasmblock}
and further simplified into
\begin{qasmblock}
inv @ z q[0];
\end{qasmblock}
without solving the problem of synthesizing \qasmline{s}. In general, however, compiling \qasmline{pow(r) @ g} involves solving a circuit synthesis or optimization problem, even in the integral case.

\subsection{Non-unitary operations}\label{sec:operations}

OpenQASM\,3 includes two basic non-unitary operations, which are the same as those found in OpenQASM\,2.

The statement \qasmline{bit = measure q} measures a qubit \qasmline{q} in the $Z$-basis, and assigns the measurement outcome, `0' or `1', to the target bit variable (see the subsection on classical types in Section~\ref{sec:rt-classical}). Measurement is `non-destructive', in that it corresponds to a \emph{projection} onto one of the eigenstates of $Z$; the qubit remains available afterwards for further quantum computation.
The \qasmline{measure} statement also works with an array of qubits, performing a measurement on each of them and storing the outcomes into an array of bits of the same length.
For example, the example below initializes, flips, and measures a register of 10 qubits.%
\begin{qasmblock}
qubit[10] qubits;
bit[10] bits;
reset qubits;
x qubits;
bits = measure qubits;
\end{qasmblock}
\noindent
For compatibility with OpenQASM\,2, we also support the syntax \qasmline{measure q -> r} for a qubit \qasmline{q} (or quantum register of some length $n$), and a bit variable \qasmline{r} (or a classical bit register of the same length). 

The statement \qasmline{reset q} resets a qubit \qasmline{q} to the state $|0\rangle$.
With idealised quantum hardware, this is equivalent to measuring \qasmline{q} in the standard basis, performing a Pauli $X$ operation on the qubit if the outcome is `1', and then discarding the measurement outcome.
Mathematically, it corresponds to a partial trace over \qasmline{q} (\emph{i.e.}, discarding it) before replacing it with a new qubit in the state $|0\rangle$.
The \qasmline{reset} statement also works with an array of qubits, resetting
them to the state $|0\rangle\otimes \cdots \otimes |0\rangle$.

\subsection{Real-time classical computing}
\label{sec:rt-classical}

OpenQASM\,2 primarily described \emph{static} circuits in which the only mechanism for control flow were \qasmline{if} statements that controlled execution of a single \qasmline{gate}. This constraint was largely imposed by corresponding limitations in control hardware. \emph{Dynamic} circuits --- with classical control flow and concurrent classical computation --- represent a richer model of computation, which includes features that are necessary for eventual fault-tolerant quantum computers that must interact with real-time decoding logic. In the near term, these circuits also allow for experimentation with qubit re-use, teleportation, and iterative algorithms.

Dynamic circuits have motivated significant advances in control hardware capable of moving and acting upon real-time values \cite{Ryan2017, Fu2017, Butko2019}.
To take advantage of these advances in control, we extend OpenQASM with classical data types, arithmetic and logical instructions to manipulate the data, and control flow keywords. While our approach to selecting classical instructions was conservative, these extensions make OpenQASM Turing-complete for classical computations in principle, augmenting the prior capability to describe static circuits composed of unitary gates and measurement.

\subsubsection*{Classical types:}

In considering what classical instructions to add, we felt it important to be able to describe arithmetic associated with looping constructs and the bit manipulations necessary to shuttle data back and forth between qubit measurements and other classical registers.
As a result, simple classical computations can be directly embedded within an OpenQASM\,3 circuit.
To support these, we introduce classical data types like signed/unsigned integers and floating point
values, to provide well-defined semantics for arithmetic operations.

The OpenQASM type system was designed with two distinct requirements in mind.
When used within high-level logical OpenQASM circuits the type system must capture the programmer's intent. It must also remain portable to different kinds of quantum computer controllers. When used within low-level physical OpenQASM
circuits the type system must reflect the realities of particular controller, such as limited memory
and well-defined register lengths.
OpenQASM takes inspiration from type systems within classical programming languages but adds some
unique features for dynamic quantum circuits.

For logical-level OpenQASM circuits, we introduce some
common types which will be familiar to most programmers: \qasmline{int} for signed integers,
\qasmline{uint} for unsigned integers, \qasmline{float} for floating point numbers, \qasmline{bool} for boolean true or false values, and \qasmline{bit} for individual bits.
The precision of the integer and floating-point types is not specified within the OpenQASM language itself: rather the precision is presumed to be a feature of a particular controller.
Simple mathematical operations are available on these types, for example:
\begin{qasmblock}
int x = 4;
int y = 2 ** x; // 16
\end{qasmblock}
\noindent
More complex classical operations are delegated to \qasmline{extern} functions, as we describe below.

For low-level platform-specific OpenQASM circuits, we introduce syntax for defining arbitrary bit widths of classical types.
In addition to the hardware-agnostic \qasmline{int} type OpenQASM\,3 also allows programmers to specify an integer with exact precision \qasmline{int[n]} for any positive integer
\qasmline{n}. For example,

\begin{qasmblock}
uint[8] x = 0;
// some time later...
bit a = x[2]; // read the 3rd least-significant bit of 'x' and assign it to 'a'
x[7] = 1;     // update the most-significant bit of 'x' to 1
\end{qasmblock}

\noindent
shows direct access to the underlying bits in classical values. This kind of flexibility is common in hardware description languages like VHDL and Verilog where manipulations of individual bits within registers are common.
We expect similar kinds of manipulations to be common in OpenQASM\,3, and so we allow for direct access to bit-level updates within classical values.
In practice, we expect target platforms for OpenQASM\,3 to only  support particular bit widths of each classical type: \emph{e.g.},~some target hardware might only
allow for \qasmline{uint[8]} and \qasmline{uint[16]}, while other targets might only offer
\qasmline{uint[20]}.
This is intended to allow OpenQASM\,3 to describe platform-specific code which is tailored to the registers and other resources of controllers of a quantum circuit unit.
If the programmer does not require a particular bit width, they may leave the bit-width unspecified: bit-level operations on the variable would no longer be permitted, but an appropriate choice of bit-width would then be deferred to a platform-specific compiler.

OpenQASM\,3 also introduces a fixed-point
\qasmline{angle} type to represent angles or phases.
These may also be defined to have a specific bit-width through the types \qasmline{angle[n]}.
Values stored in variables of the type \qasmline{angle[n]} may be interpreted as representing values in the range $[0, 2\pi)$,\footnote{%
    The \qasmline{angle} types do not actually commit to the value of the angle being positive, and can also be interpreted as representing values in the range $[-\pi, \pi)$; the description above is provided for the sake of concreteness.
}
with a fractional binary expansion of the form $2\pi \times  0.c_{n-1}c_{n-2} \dotsm c_{0}$\,.
(For example, with an \qasmline{angle[4]}, the binary
representation of $\pi$ would be $c_3 c_2 c_1 c_0 = \texttt{1000}$, using a big-endian representation; the representation of
$\pi/2$ would be \texttt{0100}; and so forth.)
This representation of phases is common to both quantum phase estimation \cite{kitaev95} as well as numerically-controlled oscillators.
The latter case is particularly relevant to real-time control
systems that need to track qubit frames at run-time, where they can take advantage of numerical
overflow to automatically constrain the phase representation to the domain $[0, 2\pi)$, obviating
the need for expensive floating-point modular arithmetic. The confluence of this angle
representation in controllers and quantum algorithms influenced us to add first-class support for
this type in OpenQASM\,3.

\subsubsection*{Classical control flow:}

OpenQASM\,2 circuits were effectively limited to straight line code: any quantum algorithm with more involved constructs, such as loops, could only be specified through meta-programming in a more general tool such as Python. The resulting OpenQASM\,2 circuits could end up being quite lengthy in some
cases, and lost all the structure from their original synthesis.
In OpenQASM\,3, we introduce introduce \qasmline{while} loops and \qasmline{for} loops.
We also extend the
\qasmline{if} statement syntax to allow for multiple instructions within the body of the \qasmline{if}\,, and to
allow for an \qasmline{else} block to accompany it.

As an example, consider an inverse quantum Fourier
transform circuit as written in OpenQASM\,2 (Figure~\ref{fig:iqft1}):%

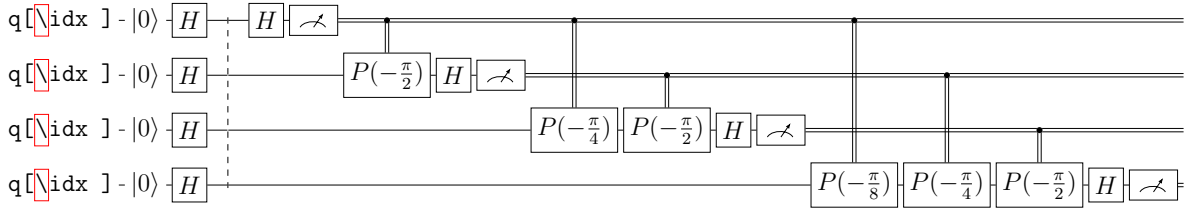
\begin{figure*}
    \centering
    \scalebox{0.8}{
        \begin{tikzpicture}
    \hspace{-5mm}
  \begin{yquant*}
      qubit {\qasmline{\reg[\idx]}\!} q[4];
      [draw=none] box {$\ket{0}$} q;
      h q;
      barrier (q[0], q[1], q[2], q[3]);
      h q[0];
      measure q[0];
      box {$P(-\frac{\pi}{2})$} q[1] | q[0];
      h q[1];
      measure q[1];
      box {$P(-\frac{\pi}{4})$} q[2] | q[0];
      box {$P(-\frac{\pi}{2})$} q[2] | q[1];
      h q[2];
      measure q[2];
      box {$P(-\frac{\pi}{8})$} q[3] | q[0];
      box {$P(-\frac{\pi}{4})$} q[3] | q[1];
      box {$P(-\frac{\pi}{2})$} q[3] | q[2];
      h q[3];
      measure q[3];
  \end{yquant*}
        \end{tikzpicture}
        } % end scalebox
        \caption{Fourier sampling circuit with the loop unrolled.}
        \label{fig:iqft1}
\end{figure*}

\begin{qasmblock}
// Fourier basis sampling procedure, with the loop unrolled
qreg q[4];
creg c[4];

// An example preparation of register q, to be sampled:
reset q;
h q;
barrier q;

// Fourier sampling circuit:
h q[0];
measure q[0] -> c[0];

if (c[0] == 1) p(-π/2) q[1];
h q[1];
measure q[1] -> c[1];

if (c[0] == 1) p(-π/4) q[2];
if (c[1] == 1) p(-π/2) q[2];
h q[2];
measure q[2] -> c[2];

if (c[0] == 1) p(-π/8) q[3];
if (c[1] == 1) p(-π/4) q[3];
if (c[2] == 1) p(-π/2) q[3];
h q[3];
measure q[3] -> c[3];
\end{qasmblock}
\noindent
(The \qasmline{barrier} statement is discussed in Section~\ref{sec:timing}.)
One can recognize a natural iterative structure to this circuit, but the lack of a looping construct
has forced us to unroll this structure when writing it out. Furthermore, each iteration has a series
of \qasmline{if} statements that in principle could be combined, as the body of those statements is
a $Z$ rotation in all cases.
In OpenQASM\,3, we might instead write the above circuit as follows:%

\begin{qasmblock}
// Fourier sampling circuit, expressed with a for loop
qubit[4] q;
bit b;
angle[4] θ = 0; // an angular parameter, for classically controlled Z rotations

// An example preparation of register q, to be sampled:
reset q;
h q;
barrier q;

// Fourier sampling circuit:
for i in [0:3] {
    θ >>= 1;        // Divide by 2, all the rotation angles controlled by past measurement outcomes
    p(-θ) q[i];     // Do the classically controlled Z rotations all at once
    h q[i];
    measure q[i] -> b;
    θ[3] = b;       // Set new highest bit of θ to measurement outcome
}
\end{qasmblock}

\noindent
which preserves the iterative structure of the inverse QFT and uses the \qasmline{angle[4]} type (and operations on its bit-level representation) to directly convert measurement bits into the appropriate Z rotations.

\subsubsection*{Subroutines:}

Sufficiently simple functions (accompanied by sufficiently capable controllers) allow for classical subroutines to be fully implemented within the language itself.
Consider a simple \qasmline{vote} function which takes the majority `vote' of three \qasmline{bit} parameters.
This could be implemented as follows:
\begin{qasmblock}
def vote(bit a, bit b, bit c) -> bit {
    int count = 0;
    if (a == 1) count++;
    if (b == 1) count++;
    if (c == 1) count++;

    if (count >= 2) {
        return 1;
    } else {
        return 0;
    }
}
\end{qasmblock}

We extend this functionality to quantum procedures as well.
For instance, a repeat-until-success quantum circuit~\cite{rus} is naturally
described by a \qasmline{while} loop involving measurements, where the computation terminates only after a particular measurement outcome is observed. That is, the classical controllers make dynamic decisions about which operations are to be performed.
The following is one example:

\begin{qasmblock}
// Repeat-until-success circuit for rz(π + θ) where cos(θ)=3/5

// This subroutine applies the identity to psi if the output is 01, 10, or 11.
// If the output is 00, it applies a Z-rotation by the angle θ + π where cos(θ)=3/5.
def segment(qubit[2] anc, qubit psi) -> bit[2] {
  bit[2] b;
  reset anc;
  h anc;
  ccx anc[0], anc[1], psi;
  s psi;
  ccx anc[0], anc[1], psi;
  z psi;
  h anc;
  b = measure anc;
  return b;
}

// Main circuit for the repeat-until-success circuit
qubit in;
qubit[2] ancilla;
bit[2] flags = b"11";
bit out;

reset in;
h in;

// The segment subroutine returns the desired outcome 00 with probability 5/8,
// so we iterate a non-deterministic number of times until that outcome is seen.
while (flags != b"00") {       // braces are optional in this case
  flags = segment(ancilla, in);
}
rz(π - arccos(3 / 5)) in;  // total rotation of 2*π
h in;
out = measure in;              // output should equal zero
\end{qasmblock}

\subsubsection*{External functions:}

Rather than bulk up the feature set of OpenQASM for classical computation further --- which would require sophisticated classical compiler infrastructure to manage --- we selected only those elements we thought likely to be used frequently.
To provide access to more sophisticated classical computations, we instead introduce a \qasmline{extern} mechanism to connect OpenQASM\,3 circuits to arbitrary, opaque classical computations.

An \qasmline{extern} is declared similarly to a function declaration in a C header file. That is, rather than define commonly used simple subroutines in OpenQASM, the programmer may provide a signature of the form:

\begin{qasmblock}
extern vote(bit a, bit b, bit c) -> bit;
\end{qasmblock}

\noindent
and then use \qasmline{vote} like a subroutine acting on classical \qasmline{bits}.
In order to connect to existing classical compiler infrastructure, the definition of an \qasmline{extern} is not embedded within an OpenQASM circuit, but is expected to be provided to the compilation tool chain in some other language format such as Python, C, x86 assembly, or LLVM IR. This allows externs to be compiled with tools like GCC or LLVM with a minimal amount of additional structure to manage execution within the run-time requirements of a global controller. Importantly, this strategy of connecting to existing classical programming infrastructure implies that the programmer can use existing libraries without porting the underlying code into a new programming language.

Unlike other approaches to interfacing with classical computing, OpenQASM does not have an explicit synchronization primitive such as the WAIT instruction in QUIL~\cite{Quil, smith16}. In particular, invoking an \qasmline{extern} function does not imply a synchronization boundary at the call site\footnote{It is possible to achieve non-blocking semantics in QUIL using memory movement instructions to invoke function execution, and then reserving WAIT for synchronization upon completion.}. A compiler is free to use dataflow analysis to insert any required target-specific synchronization primitives at the point where outputs from extern call are used. That is, invoking an \qasmline{extern} function \emph{schedules} a classical computation, but does not wait for that calculation to terminate. The \qasmline{extern} semantics abstract away the particulars of a vendor-specific application binary interface (ABI) for how data is moved to and from the classical function.

\subsection{Input and output parameters}
\label{sec:parameterised-circuits}

In addition to the real-time classical compute constructs of OpenQASM\,3, we introduce features targeted to support  near-time computation, in the form of OpenQASM programs with input and output parameters (described elsewhere as `parameterized circuits'~\cite{Karalekas_2020}).
This functionality is provided through keywords \qasmline{input} and \qasmline{output}, which act as modifiers on variable declarations, and allow OpenQASM\,3 circuits to represent these variables as accepting input parameters and returning select output parameters.

The \qasmline{input} modifier can be used to indicate that one or more variable declarations represent values which will be provided at run-time, upon invocation.
This allows the programmer to use the same compiled circuits which only differ in the values of certain parameters.
As OpenQASM\,2 did not allow for input parameters or \qasmline{input} declaration, for compatibility OpenQASM\,3 does not require an \qasmline{input} declaration to be provided: in this case it assumes that there are no input parameters.
When an \qasmline{input} declaration is provided, the compiler produces an executable that leaves these free parameters unspecified: a circuit run would take as input both the executable and some choice of the parameters.

The \qasmline{output} modifier can be used to indicate that one or more variables are to be provided as an explicit output of the quantum procedure.
Note that OpenQASM\,2 did not allow the programmer to specify that only a subset of its variables should be returned as output, and so it would return all classical variables (which were all \qasmline{creg} variables) as output.
For compatibility, OpenQASM\,3 does not require an \qasmline{output} declaration to be provided: in this case it assumes that all of the declared variables are to be returned as output.
If the programmer provides one or more \qasmline{output} declarations, then only those variables described as outputs will be returned as an output of the quantum process.
A variable may not be marked as both \qasmline{input} and \qasmline{output}.

The \qasmline{input} and \qasmline{output} modifiers allow the programmer to more easily write variational quantum algorithms: a quantum algorithm with some free parameters, which may be run many times with different parameter values which are determined by a classical optimiser at near-time.
Rather than write a circuit which generates a new sequence of  operations for each run, OpenQASM\,3 allows such circuits to be expressed as a single program with input parameters.
This allows the programmer to communicate many different circuits with a single file, which only has to be compiled once, amortizing the cost of compilation across many runs.
For an example, we may consider a circuit which performs a measurement in a basis given by an input parameter:

\begin{qasmblock}
input int basis;  // 0 = X basis, 1 = Y basis, 2 = Z basis
output bit result;
qubit q;

// Some complicated circuit...

if (basis == 0) h q;
else if (basis == 1) rx(π/2) q;
result = measure q;
\end{qasmblock}

\noindent

\noindent
For a second example, consider the Variable Quantum Eigensolver (VQE)
algorithm~\cite{vqe}.
In this algorithm the same circuit is repeated many times using
different sets of free parameters to minimize an expectation value.
The following is an example, in which there is also more than one input variable:
\begin{qasmblock}
input angle param1;
input angle param2;
qubit q;

// Build an ansatz using the above free parameters, eg.
rx(param1) q;
ry(param2) q;

// Estimate the expectation value and store in an output variable
\end{qasmblock}

\noindent
The following Python pseudocode illustrates the differences between using and not using input and output parameters in a quantum program for the case of the VQE. This pseudocode is fictional but demonstrates the near-term execution model: where there is some classical code coordinating the compilation and execution of OpenQASM programs.

When input and output parameters are not used, the program must be recompiled to account for every change of the input variable theta. Compilation is assumed to be expensive and slow so it will be the main bottleneck towards completion of the overall VQE algorithm.

\begin{pythonblock}
for theta in thetas:
    # Create an OpenQASM circuit with θ defined
    circuit = subsitute_theta(read("circuit.qasm"))

    # The slow compilation step is run on each iteration of the inner loop
    binary = compile_qasm(circuit)
    result = run_program(binary)
\end{pythonblock}

When input and output parameters circuits are used, we can hoist compilation out of the loop. The only part of the program that is changing is the value of \qasmline{θ}. The entire program doesn’t need to be re-compiled each time.

\begin{pythonblock}
# io_circuit.qasm begins with the line "input angle θ;"
circuit = read("io_circuit.qasm")
# The slow compilation step only happens once
binary = compile_qasm(circuit)

for theta in thetas:
    # Each iteration of the inner loop is reduced to only running the circuit
    result = run_program(binary, θ=theta)
\end{pythonblock}

\section{Concepts of the language: the physical level}
\label{sec:physical-features}

In addition to support for logical-level quantum computations, OpenQASM\,3 has added support for  quantum computing experiments and platform-dependent tuning of quantum instructions on the physical level. At the physical level, quantum operations are implemented by causing time-varying stimuli to be transmitted to the qubits, and capturing the time-dependent responses. For example, for superconducting transmon platforms this might take the form of shaped microwave pulses transmitted via a number of coaxial cables, and for trapped ions might take the form of modulated laser pulses. There are various tradeoffs involved in optimal selection of the mapping from logical quantum operations to specific pulse implementations.  Therefore there can be benefits in allowing a programmer who has knowledge of both the hardware limitations and the algorithm requirements to influence the physical implementation for key parts of the circuit. For example, choosing a gate implementation with shorter duration but less-accurate calibration might be a good choice for a variational ansatz where the specific unitary implemented is less important than minimizing decoherence. Another example could be choosing a specific ratio of delays for the idle periods of a qubit to implement spin-echo (and more generally, dynamical decoupling) sequences to cancel out errors caused by hardware parameters that are known to drift slowly compared to the timescale of circuit execution.   

In this section we give a overview of the important features of the physical, lower level of OpenQASM\,3.
For a detailed description we recommend reading the live specification \cite{livedoc}.

\subsection{Timing and optimization}
\label{sec:timing}

A key aspect of expressing code for quantum experiments is the ability
to control the timing of gates and pulses. Examples include characterization of decoherence and
crosstalk~\cite{srb}, dynamical decoupling~\cite{hahn, cpmg, uhrig}, dynamically corrected
gates~\cite{dcg, de}, and gate parallelism or scheduling~\cite{xtalk}.
We introduce such timing semantics in OpenQASM\,3 to enable such circuits.

\subsubsection*{Delay statements and duration types:}

We introduce a \qasmline{delay} statement, which allows the programmer to specify relative timing of operations.
These statements can take the form \qasmline{delay[t]}, where \qasmline{t} is a value of type \qasmline{duration}.
We introduce the \qasmline{duration} type to support such timing instructions: values of type \qasmline{duration} represent amounts of time measured in seconds (typically also modified by an SI prefix).
For example, a qubit's relaxation time ($T_1$) can be measured by an OpenQASM snippet of
the following form:

\begin{qasmblock}
duration stride = 1us;   // a time duration, specified in SI units
int p = 3;                 // record the 3rd datapoint in the relaxation curve
reset q[0];
x q[0];
delay[p * stride] q[0];
c0 = measure q[0];       // survival probability of the excited state
\end{qasmblock}

\noindent
Furthermore, any instruction may take a bracketed duration, such as \qasmline{x[30ns] q[1]} for a gate \qasmline{x} defined with a \qasmline{gate} declaration. Instructions with bracketed durations are required to have exactly that duration (although, see the \qasmline{stretch} type below).
If an instruction is not defined by a \qasmline{defcal} consistent with the requested duration, then it may not be executable on hardware.

\subsubsection*{Boxes and barrier statements:}

OpenQASM\,3 includes features to constrain the reordering of gates, where the timing of those gates might otherwise be changed by the compiler (or the gates removed entirely as a valid optimization on the logical level).

We introduce a \qasmline{box} statement, to help with scoping the timing of a particular part of the circuit. A boxed sub-circuit is different from a \qasmline{gate} (see Section~\ref{sec:gates}) or \qasmline{def} subroutine (see Section~\ref{sec:rt-classical}), in that it is merely a pointer to a piece of code within the larger scope which contains it.
This can be used to signal permissible logical-level optimizations to the compiler: optimizing operations within a \qasmline{box} definition is permitted, and optimizations that move operations from one side to the other side of a box are permitted, but moving operations either into or out of the box as part of an optimization is forbidden.
The compiler can also be given a description of the operation which a \qasmline{box} definition is meant to realise, allowing it to re-order gates \emph{around} the box.
For example, consider a dynamical decoupling sequence inserted in a part of the circuit:
\begin{qasmblock}
rx(2*π/12) q;
box {
    delay[ddt] q;
    x q;
    delay[ddt] q;
    x q;
    delay[ddt] q;
}
rx(3*π/12) q;
\end{qasmblock}
By boxing the sequence, we create a box that implements the identity.
The compiler is now free to commute a gate past the box by knowing the unitary implemented by the box:
\begin{qasmblock}
rx(5*π/12) q;
box {
    delay[ddt] q;
    x q;
    delay[ddt] q;
    x q;
    delay[ddt] q;
}
\end{qasmblock}
The compiler can thus perform optimizations without interfering with the implementation of the dynamical decoupling sequence.

As with other operations, we may use square brackets to assign a duration to a box: this can be used to put hard constraints on the execution of a particular sub-circuit by requiring it to have the assigned duration.
This can be useful in scenarios where the exact duration of a piece of code is unknown (\emph{e.g.},~if it is runtime dependent), but where it would be helpful to impose a duration on it for the purposes of scheduling the larger circuit. For example, if the duration of the parameterized gates \qasmline{mygate1(a, b)}, \qasmline{mygate2(a, b)} depend on values of the variables \qasmline{a} and \qasmline{b} in a complex way, but an offline calculation has shown that the total will never require more than 150ns for all valid combinations:
\begin{qasmblock}
// some complicated circuit that gives runtime values to a, b
box [150ns] {
    delay[str1] q1; // schedule as late as possible within the box
    mygate1(a, a+b) q[0], q[1];
    mygate2(a, a-b) q[1], q[2];
    mygate1(a-b, b) q[0], q[1];
}
\end{qasmblock}
In the above example \qasmline{str1} is of \qasmline{stretch} type, to be discussed in the next section, and in this the compiler will resolve it to the proper value to fill out the box to \qasmline{150ns}.

OpenQASM\,3 also retains the \qasmline{barrier} instruction of OpenQASM\,2, which prevents gates from being reordered across its source line, either on a specified qubit (or register) \qasmline{q}, or a sequence of qubits (or registers) \qasmline{q, r, ...} \,.
For example,

\begin{qasmblock}
cx r[0], r[1];
h q[0];
h p[0];
barrier r, q[0];
h p[0];
cx r[1], r[0];
cx r[0], r[1];
\end{qasmblock}

\noindent
This will prevent the compiler from performing optimizations which involve combining or canceling the CNOT gates on the register \qasmline{r}.
However, it does not prevent similar optimizations of the \qasmline{h p[0]} operations on either side of it.
Outside of any \qasmline{box}, a \qasmline{barrier} statement may also be invoked without any arguments (as in the examples of classical control flow in Section~\ref{sec:rt-classical}), in which case it prevents gates from being reordered across its source line, on \emph{all} qubits.

More generally, \qasmline{delay} statements also prevent optimizations or re-orderings of gates, which would move a gate acting on some qubit past a \qasmline{delay} statement on the same qubit.
The \qasmline{barrier} instruction in OpenQASM\,3 is similar to a special case of \qasmline{delay}, with a duration of zero. The difference between \qasmline{barrier} and \qasmline{delay[0]} is that the use of \qasmline{delay} indicates a fully scheduled series of instructions, that any should not be altered by further scheduling, whereas \qasmline{barrier} merely indicates an ordering constraint, and if necessary a scheduling pass may add finite-duration delays adjacent to \qasmline{barrier}s.
In this way, \qasmline{barrier} statements may be regarded as helping to provide time synchronization within the given scope and across the qubits on which they are applied.
However, the constraints implied by \qasmline{delay} and \qasmline{barrier} statements do not leak out of any containing \qasmline{box} declarations, and thus do not prevent commutation with uses of the \qasmline{box}: they are treated as implementation details of the box, and are invisible to the outside scope.

\subsubsection*{Stretch types:}

By specifying relative timing of operations rather than absolute timing, it is possible to use OpenQASM\,3 to provides more flexible timing of operations, which can be helpful in a setting with a variety of calibrated gates with different durations.
We introduce a new type \qasmline{stretch}, representing a duration of time which is resolvable to a concrete duration at compile time once the exact duration
of calibrated gates are known.
A \qasmline{stretch} quantity is a value which the compiler attempts to minimise, subject to timing constraints on one or more qubits.
(We say that each a \qasmline{stretch} value has a `natural duration' of 0, though constraints may impose a larger value on any given \qasmline{stretch} value.)
This makes it possible for a programmer/compiler to express gate designs such as evenly spacing gates (\emph{e.g.},~to implement a
higher-order echo decoupling sequence),  left-aligning a sequence of gates, or applying a gate for the duration of some subcircuit.
We are motivated to include this feature by the importance of specifying gate timing and parallelism, in a way that is independent of the precise
duration and implementation of gates at the pulse-level description.
This increases circuit portability by decoupling circuit timing
intent from the underlying pulses, which may change from machine to machine or even from day to day.

We refer to the problem of assigning definite values to \qasmline{stretch} values, as ``the stretch problem''.
The \qasmline{stretch} type itself, and our approach to the stretch problem, is inspired by how a similar problems are solved in \TeX~\cite{knuth84} using ``glues'', where the compiler spaces out letters, words, etc. appropriately given a target font.
In quantum circuits we have the additional challenge that timings on different qubits are not independent, \emph{e.g.},~if a two-qubit gate connects the qubits.
This means that the choice of timing on one qubit line can have a ripple effect on other qubits.
We describe how the stretch problem is solved through multiobjective linear programming in Section~\ref{sec:stretch-problem}.

\begin{figure*}
    \centering
    \begin{subfigure}[b]{\textwidth}
    \centering
    \scalebox{.9}{
        \begin{tikzpicture}
  \begin{yquant}
      qubit {\qasmline{\reg[\idx]}\!} q[5];
      cnot q[1] | q[0];
      box {$U(\frac{\pi}{4}, 0, \frac{\pi}{2})$} q[2];
      cnot q[3] | q[4];
      ["north:g1" {font=\protect\tiny, inner sep=0pt}", name=boxa, draw=none] box {\hspace{52pt}} (q[0], q[1]);
      ["north:g2" {font=\protect\tiny, inner sep=0pt}", name=boxb, draw=none] box {\hspace{12pt}} q[2];
      ["north:g3" {font=\protect\tiny, inner sep=0pt}", name=boxc, draw=none] box {\hspace{52pt}} (q[3], q[4]);
  \end{yquant}
    \spring{boxa}{4mm};
    \spring{boxb}{2mm};
    \spring{boxc}{4mm};
        \end{tikzpicture}
        }
        \label{fig:leftalign}
    \hfill
    \centering
        \scalebox{.9}{
        \begin{tikzpicture}
  \begin{yquant}
      qubit {\texttt{\$\idx}} q[5];
      [x radius=38.5mm] box {$CX [300 ns]$} (q[0], q[1]);
      [x radius=6.5mm] box {$U [50 ns]$} q[2];
      [x radius=25.5mm] box {$CX [200 ns]$} (q[3], q[4]);
      [x radius=31mm] box {$Delay [250 ns]$} q[2];
      [x radius=12.5mm] box {$Delay [100 ns]$} (q[3], q[4]);
  \end{yquant}
        \end{tikzpicture}
        }
        \label{fig:leftalignresolved}
        \caption{}
    \end{subfigure}
	\par\bigskip
	\par\bigskip
    \begin{subfigure}[b]{\textwidth}
    \centering
        \scalebox{.9}{
        \begin{tikzpicture}
    \begin{yquant}
      qubit {\qasmline{\reg[\idx]}\!} q[3];
      cnot q[1] | q[0];
      ["north:g" {font=\protect\tiny, inner sep=0pt}", name=boxa, draw=none] box {\hspace{9pt}} q[2];
      box {$U(\frac{\pi}{4}, 0, \frac{\pi}{2})$} q[2];
      ["north:2*g" {font=\protect\tiny, inner sep=0pt}", name=boxb, draw=none] box {\hspace{18pt}} q[2];
    \end{yquant}
    \spring{boxa}{2mm};
    \spring{boxb}{2mm};
        \end{tikzpicture}
        }
        \label{fig:thirdalign}
    \hfill
    \centering
        \scalebox{.9}{
        \begin{tikzpicture}
    \begin{yquant}
      qubit {\texttt{\$\idx}} q[3];
      [x radius=46mm] box {$CX [350 ns]$} (q[0], q[1]);
      [x radius=13mm] box {$Delay [100 ns]$} q[2];
      [x radius=6mm] box {$U [50 ns]$} q[2];
      [x radius=25mm] box {$Delay [200 ns]$} q[2];
    \end{yquant}
        \end{tikzpicture}
        }
        \label{fig:thirdalignresolved}
        \caption{}
    \end{subfigure}
    \caption{Arbitrary alignment of gates in time using stretchy delays.
    a) left-justified alignment intent, and timed instructions after stretches are resolved.
    b) alignment intent of a short gate at the one-third point of a long gate, and after stretch resolution.
    \label{fig:alignment}}
\end{figure*}
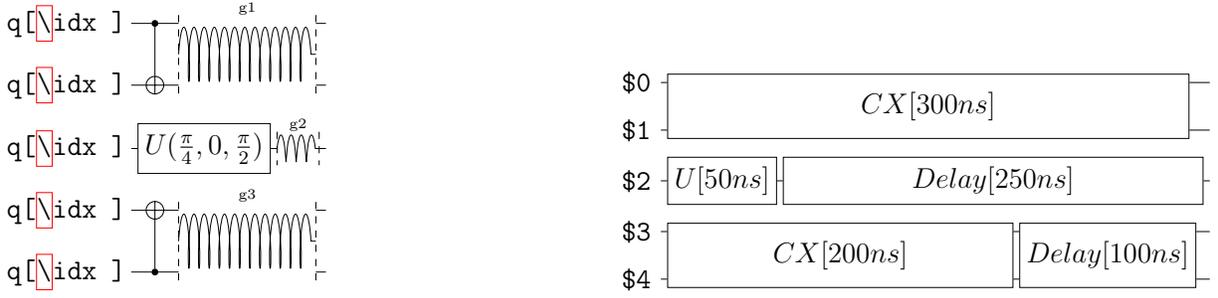
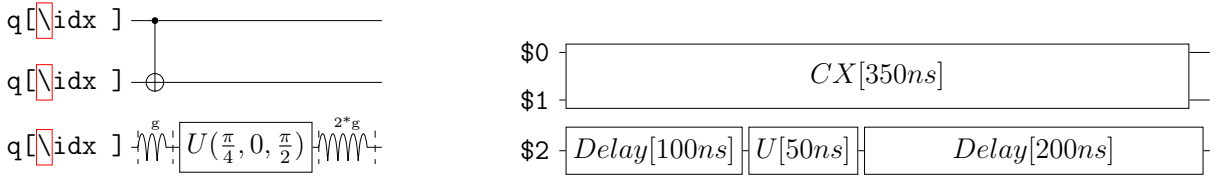

A simple use case for stretchable delays are gate alignments. We can use this feature to gain
control on how gates are aligned in a circuit, regardless of the latencies of those gates on the machine.
For example, consider the following code sample for left-aligning operations on a collection of qubits:

\begin{qasmblock}
// left alignment
stretch g1, g2, g3;
barrier q;
cx q[0], q[1];
delay[g1] q[0], q[1];
u(π/4, 0, π/2) q[2];
delay[g2] q[2];
cx q[3], q[4];
delay[g3] q[3], q[4];
barrier q;
\end{qasmblock}
\noindent
The corresponding circuit is illustrated in
Figure~\ref{fig:alignment}a (where the spring symbols on and across qubit wires denote delays with \qasmline{stretch} parameters).
After the circuit is lowered to a particular machine with known gate calibrations and durations,
the compiler resolves every stretchy \qasmline{delay} instruction into one with concrete timing, so that the operations performed on each qubit take the same cumulative amount of time.

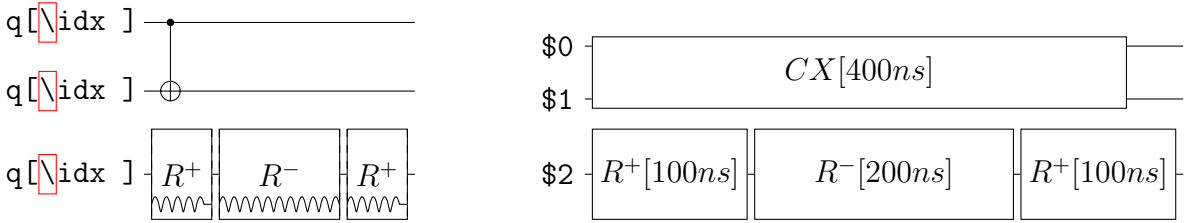
\begin{figure*}
    \centering
    \begin{subfigure}[b]{.35\textwidth}
        \centering
        \begin{tikzpicture}
           \begin{yquant}
              qubit {\qasmline{\reg[\idx]}\!} q[3];
              cnot q[1] | q[0];
              [name=boxa, x radius=4mm, y radius=6mm] box {$R^+$} q[2];
              [name=boxb, x radius=8mm, y radius=6mm] box {$R^-$} q[2];
              [name=boxc, x radius=4mm, y radius=6mm] box {$R^+$} q[2];
           \end{yquant}
            \spring[4mm]{boxa}{1mm};
            \spring[4mm]{boxb}{1mm};
            \spring[4mm]{boxc}{1mm};
        \end{tikzpicture}
    \end{subfigure}
    \hspace*{\fill}
    \begin{subfigure}[b]{.6\textwidth}
        \centering
    \begin{tikzpicture}
       \begin{yquant}
          qubit {\texttt{\$\idx}} q[3];
          [x radius=35.5mm] box {$CX [400 ns]$} (q[0], q[1]);
          [x radius=8.6mm, y radius=6mm] box {$R^+ [100 ns]$} q[2];
          [x radius=17.2mm, y radius=6mm] box {$R^- [200 ns]$} q[2];
          [x radius=8.6mm, y radius=6mm] box {$R^+ [100 ns]$} q[2];
       \end{yquant}
    \end{tikzpicture}
    \end{subfigure}
    \caption{Dynamically corrected CNOT gate where the spectator has a rotary pulse. The rotary gates are stretchy, and the design intent is to interleave a "winding" and "unwinding" that is equal to the total duration of the CNOT. We do this without knowledge of the CNOT duration, and the compiler resolves them to the correct duration during lowering to the target backend.}
    \label{fig:dcg}
\end{figure*}

We can further control the exact alignment by giving relative weights to the stretchy delays.
This method is flexible, e.g. to align a gate at the 1/3 point of another gate, we can use
a \qasmline{delay[g]} instruction before, and a \qasmline{delay[2*g]} after the gate to be aligned, as in the following example (illustrated as a circuit in Figure~\ref{fig:alignment}b):

\begin{qasmblock}
// one-third alignment
stretch g;
barrier q;
cx q[0], q[1];
delay[g] q[2];
u(π/4, 0, π/2) q[2];
delay[2*g] q[2];
barrier q;
\end{qasmblock}

The scope of a \qasmline{stretch} variable can be bounded by a \qasmline{box} statement.
If the \qasmline{box} statement has been assigned a duration (\emph{e.g.},~to help with scheduling operations in the larger circuit), this imposes a limitation on the total duration of the operations to be performed within the \qasmline{box}, which in turn constrains the \qasmline{stretch} values inside.
For example:
\begin{qasmblock}
// define a 1ms box whose content is just a centered CNOT
box [1ms] {
    stretch a;
    delay[a] q;
    cx q[0], q[1];
    delay[a] q;
}
\end{qasmblock}
\noindent
The `natural' duration of instructions within a box (\emph{e.g.},~taking a hypothetical value of 0 for all stretch values) must be smaller than the declared box duration, otherwise a compile-time error will be raised.
The \qasmline{stretch} inside the box will always be set to fill the difference between the declared duration and the natural duration.
If the instructions exceed the deadline at runtime, a runtime error is raised.

Instructions other than \qasmline{delay} can also be `stretchy', if they are explicitly defined as such.
Consider for example a rotation called \qasmline{rotary} that is applied for the entire duration
of some other gate~\cite{pong}. Once resolved to a concrete time, it can be passed to
the gate definition (in terms of pulses) to realize the gate.

\begin{qasmblock}
// stretchy rotation gate
stretch g;
box {
    cx control, target;
    rotary(amp)[g] spectator;    // rotate for some duration, with some pulse amplitude
    rotary(-amp)[2*g] spectator; // rotate back, for twice that duration
    rotary(amp)[g] spectator;    // rotate forward again, making overall identity
}
\end{qasmblock}

\noindent
In particular, the duration of a \qasmline{box} statement can also be be given as a \qasmline{stretch}, in which case its `natural' duration is constrained by the natural durations of its operations, and the duration it takes at run-time depends on the constraints imposed in the larger circuit in which the \qasmline{box} block occurs.
For example,
\begin{qasmblock}
// using box to label a sub-circuit, and later delaying for that entire duration
stretch mybox;
box [mybox] {
    cx q[0], q[1];
    delay[200ns] q[0];
}
delay[mybox] q[2], q[3];
cx q[2], q[3];
\end{qasmblock}
\noindent
realises a CNOT operation on qubits \qasmline{q[0]} and \qasmline{q[1]}, followed by a delay of at least 200ns, followed by a CNOT  on qubits \qasmline{q[2]} and \qasmline{q[3]}.

Because stretches must be resolved through solving a linear program, there are limitations on their use: they can appear only in linear combinations of stretch variables and durations, and their numerical value is only available at a late stage in the compilation process. Because \qasmline{duration} types are known explicitly without solving a linear program, they can be used more freely in nonlinear expressions and control flow. The following example is valid with \qasmline{pi_time} as \qasmline{duration} but would be invalid if it were a \qasmline{stretch}.
\begin{qasmblock}
// sequence of 30ns π pulses to fill total duration
duration total_time = 1000ns;
duration pi_time = 30ns;
for _ in [1:floor(total_time/pi_time)] {
    x[pi_time] q[0];
}
\end{qasmblock}

\subsection{Calibrating quantum operations}\label{sec:calibrating}

OpenQASM describes circuits applying quantum operations such as unitary gates and projective measurements to qubits. Quantum operations are typically implemented with classical time-dependent signals coupled to quantum systems. Control hardware such as arbitrary waveform generators, flux sources, and lasers emit signals to orchestrate the synchronous emission of calibrated control fields. These guide the quantum system to implement the desired quantum operation~\cite{Glaser2015}.
Consequently, control system implementations at the hardware and software level are highly platform-dependent. For example, superconducting transmon qubits encode a qubit in a non-linear oscillator formed by a parallel circuit consisting of a Josephson junction and a capacitor. To manipulate the state of the qubit a series of shaped microwave control pulses are applied to it~\cite{Koch2007}. The design of these control signals is an active area of research, and developments in this area enable software updates to improve the performance of existing hardware systems~\cite{jurcevic2021demonstration}. The support for calibration grammars in OpenQASM is motivated by the rapid pace of development within the quantum hardware space and the need for hardware/software co-design to extract the maximum performance from today's quantum computers. Similar considerations have motivated several software projects to directly support signal-level control of quantum hardware, including QGL~\cite{QGL}, Artiq~\cite{artiq}, OpenPulse~\cite{qiskitpulse, mckay2018}, JaqalPaw~\cite{jacalpaw}, QUIL-T~\cite{quilt}, QCCS~\cite{qccs}, Qua~\cite{qua}, and Pulser~\cite{pulser}.

OpenQASM has added support for specifying instruction calibrations in the form of \qasmline{defcal} (short for ``define calibration'') declarations which allow the programmer to specify a microcoded~\cite{wilkesBestWayDesign1989} implementation of a \qasmline{gate}, \qasmline{measure}, or \qasmline{reset} instruction within a lower-level control grammar as implemented by the target hardware vendor. This particular language keyword is borrowed from the recently proposed QUIL-T extension of the Quil language~\cite{quilt}. Support for \qasmline{defcal} declarations is optional for OpenQASM compilers and target devices. The example below shows how the programmer can select the \textit{openpulse} calibration grammar (introduced later in this section), and declare the calibration for an \qasmline{rx} instruction.

\begin{qasmblock}
// select the defcalgrammar to use
defcalgrammar "openpulse";

// define the rx gate
gate rx(θ) q {
    U(θ, -π/2, -π/2) q;
}

// define the implementation of the rx-gate for physical qubit 0
defcal rx(angle[20] θ) $0 { ... }

rx(π/2) $0; // the defcal is automatically linked at compile time

\end{qasmblock}

The available calibration instructions are determined by the user's selected \qasmline{defcalgrammar}. A \qasmline{defcal} declaration has a similar form to a \qasmline{gate} declaration, but with some subtle differences. Whereas a \qasmline{gate} declaration defines a previously undefined gate in terms of other OpenQASM instructions, a \qasmline{defcal} declares the implementation of an OpenQASM instruction for a target device. In a \qasmline{gate}, parameters are infinite precision angles in the \qasmline{gate}, whereas in the \qasmline{defcal} a precision may be specified, enabling the definition of gate calibrations at the native precision of the target hardware. The \qasmline{defcal} is defined for physical qubits rather than being the definition of a unitary gate applicable to any qubit. In this case the \qasmline{defcal} is for a fixed qubit (\qasmline{$0}), \emph{i.e.},~the zeroth physical qubit on the device. (OpenQASM reserves identifiers of the form \qasmline{$n}, where $n$ is a non-negative integer for physical qubits of the corresponding integer index in the target device.) We assume that compiler mapping passes assign virtual qubits to particular physical qubits in the device.

Calibration grammars may require the ability to insert top-level configuration information and system setup outside of the \qasmline{defcal} in order to share that information across multiple \qasmline{defcal}s. To facilitate this, OpenQASM\,3 introduces \qasmline{cal} blocks. Within \qasmline{cal} blocks the semantics of the selected calibration grammar are valid. Values declared within the \qasmline{cal} block may only be referenced from other \qasmline{cal} blocks or \qasmline{defcal} declarations that capture values from the containing scope. The example below uses a \qasmline{cal} block to define frequencies, ports, and frames to be used in later \qasmline{defcal} declarations.

\begin{qasmblock}
cal {
   // defined within `cal`, so it may not leak 
   // back out to the enclosing block's scope
   float qubit_freq = 5.2e9;
   // declare an external port (see below for more information)
   extern port d0;
   // declare a new frame for port d0
   frame frd0 = newframe(d0, qubit_freq, 0.0);
}
\end{qasmblock}

Through \qasmline{defcal} and \qasmline{cal} declarations, the programmer may define the implementation of canonical instructions for the target hardware.
These operation definitions are not typically invoked from a logical-level specification of an operation. Rather, a target-system compiler links the circuit-level OpenQASM circuit to these definitions, as shown in the \qasmline{rx} definition example above. This embeds an extendable translation layer between the circuit model and control-hardware implementations of quantum instructions within OpenQASM. If a \qasmline{defcal} does not have a corresponding \qasmline{gate} declaration the compiler will treat the call to this \qasmline{defcal} as a call to an opaque circuit operation of the same name and signature. In such a case, the compiler is not able to reason about the state of the circuit across calls to these \qasmline{defcal}s. Consequently, this functionality replaces the \qasmline{opaque} keyword of OpenQASM\,2.

\subsubsection*{Calibration resolution}

The signals required to enact gates on physical hardware may vary greatly based on the kind of gate, the physical qubit type, and the input gate parameters involved.
The role of \qasmline{defcal} declarations is to specify how a \qasmline{gate} or primitive operation is performed, with \qasmline{defcal} dispatch support for specialized definitions for specific physical qubits and gate parameters.

For instance, \qasmline{defcal rx(angle[20] θ) $q} defines the implementation of a \qasmline{rx} gate of any angle
on any physical qubit \qasmline{$q}, whereas \qasmline{defcal rx(π) $0} specifically targets an $X$ rotation of $\pi$ radians on physical qubit \qasmline{$0}. The reference to physical rather than virtual qubits is critical, as quantum registers are normally not interchangeable from the perspective of the details of the control stimulus that enacts them.

At compile-time, specialized \qasmline{defcal} declarations are greedily matched with qubit
specializations selected before instruction arguments. If not all parameters are specialized, the
declaration with the largest number of specialized parameters is matched first. In the event of a
tie the compiler will choose the \qasmline{defcal} block that appeared first in the circuit.\footnote{This strategy is loosely inspired by Haskell's pattern matching for resolving function calls~\cite{marlow2010haskell} and a similar strategy appears in Quil-T's \qasmline{defcal}~\cite{quilt}.}
For instance, given:
\begin{qasmblock}
defcal rx(angle[20] θ) $q  { ... }
defcal rx(angle[20] θ) $0  { ... }
defcal rx(π/2) $0 { ... }
\end{qasmblock}
\noindent
the operation \qasmline{rx(π/2) $0} would match the \qasmline{defcal} declaration on line~3;\, \qasmline{rx(π) $0} would match the declaration on line~2;\,
and \qasmline{rx(π/2) $1} would match the declaration on line~1.
Calibrations for measurements are specified similarly, apart from having a classical bit as part of its signature, \emph{e.g.},~\qasmline{defcal measure $0 -> bit}\,.

To integrate with OpenQASM's timing system when performing scheduling, the compiler requires the set of physical qubit resources used by the \qasmline{defcal} implementation as well as the duration of its execution.
The scheduler uses this information to resolve the scheduled circuit specified by the programmer.
For this reason, all resolved \qasmline{defcal} usages must have compile-time computable durations. If the duration of a \qasmline{defcal} depends on its input parameters, the parameters and consequently the definition must be resolvable at compile-time to enable deterministic scheduling of the operations at the circuit level. Calibrations must also be independent of when the operation is invoked, to allow their substitution anywhere the corresponding circuit instruction is applied. Note that some instructions, such as a \qasmline{reset} instruction implemented with measurement feedback, might require control flow. A calibration grammar should support such use cases while still guaranteeing the deterministic duration of the instruction.

\subsubsection*{Calibration grammars}

Describing the application of control fields (or other operation specifics) for various quantum platforms may be vastly different. For example, a SWAP gate within a superconducting qubit system may be actuated with a time-dependent microwave stimulus applied to the target qubits.
In contrast, an ion-trap system may implement a SWAP gate by physically exchanging two ions through the application of a series of DC potentials and laser pulses, which may be global in nature~\cite{kielpinski2002architecture, kaufmann2017fast}. The control fields, and correspondingly, the semantics required to describe these two operations may vary significantly between hardware vendors.

In OpenQASM we do not attempt to capture all such possibilities, but rather embrace the role of a pluggable multi-level intermediate representation. Vendors are free to define their calibration grammar and an accompanying implementation. OpenQASM, in turn, provides the \qasmline{defcalgrammar <grammar>} declaration which allows the programmer to inform the compiler implementation of which calibration grammar should be selected for the following calibration declarations. Calibration support is optional, and OpenQASM compiler implementations that support defining calibrations should provide an interface for new grammars to plugin and extend the compilation infrastructure. It is then the responsibility of the hardware provider to specify a grammar and implement the required hooks into the compiler. This might be as simple as a \qasmline{defcal} mapping to custom hardware calibrations within the target system, or as complex as a fully-fledged pulse-programming language as explored later in this section.

By allowing vendors to extend the calibration grammar, OpenQASM provides a standardized pipeline for vendors to lower from a hardware-agnostic representation into one that is hardware-aware. It is the responsibility of the hardware vendor to consume the calibrations along with the rest of the OpenQASM and to compile the circuit into a final executable for the target hardware.

\subsubsection*{OpenPulse grammar}
\label{sec:openpulse-grammar}
OpenPulse was initially specified as a RESTful API system that was later extended with an implementation for Qiskit \cite{mckay2018, qiskitpulse}. Within the OpenQASM specification, the OpenPulse grammar provides a hardware-independent representation for programming qubits at the level of microwave pulses \cite{livedoc}. In a typical quantum computing system, control electronics apply stimulus in the form of time-varying pulses to hardware ports. These enact control fields upon the target quantum system. For example, in a superconducting qubit system, smooth pulse envelopes with a carrier frequency on resonance with the superconducting qubit are applied through an arbitrary waveform generator to excite the qubit from its ground state to its excited state. OpenPulse was originally designed to target superconducting qubit systems. In general, it captures the control requirements of quantum systems defined by a time-dependent Hamiltonian of the form $H(t) = H_\mathrm{sys} + \sum_k \alpha_k(t)H_k$ where $H_\mathrm{sys}$ is a time-independent drift Hamiltonian and $H_k$ are control-fields modulated by the time-dependent control knobs $\alpha_k(t)$ described within the OpenPulse program in the form of pulses emitted to hardware ports. This system model describes many quantum technologies such as ion traps, quantum dots, and neutral atoms \cite{bruzewiczTrappedionQuantumComputing2019, zwanenburgSiliconQuantumElectronics2013, pulser}. For this reason, it is the canonical \qasmline{defcalgrammar} provided within the OpenQASM3 specification.

In OpenPulse all operations fundamentally operate on a \qasmline{port} which is a software abstraction representing any input or output component meant to manipulate and observe qubits. As they are uniquely defined by the target system they are declared as \qasmline{extern port d0;}. Ports are ultimately mapped to some combination of hardware resources. For instance, a port may directly correspond to a digital-to-analog converter. In the quantum system a port typically corresponds to a Hamiltonian term, $H_k$ with pulses applied to the port defining the time-dependent $\alpha_k(t)$ control-term. For instance, a port could apply a time-dependent voltage to a qubit, or manipulate the coupling between two neighboring qubits. Ports are both vendor and device specific.

To emit pulses that track the frame of the qubit OpenPulse introduces the \qasmline{frame} as the canonical value to which pulses are applied. It is akin to the rotating frame of a Hamiltonian and is responsible for tracking two properties of pulses emitted on a \qasmline{port}. Firstly, the \qasmline{frame} represents a carrier signal $e^{i\left(2\pi f t + \phi\right)}$, parameterized  by a frequency $f$ and phase $\phi$. Frames are tracked throughout a circuit execution by the system's runtime. Any modifications to a frame within a calibration entry will persist to calls to the same frame elsewhere in the circuit. In this way, it is analogous to a numerically controlled oscillator (NCO). One motivation for keeping track of accrued phase is to allow the natural implementation of the ``virtual Z-gate''. This does not require a physical pulse but rather shifts the phase of all future pulses on that frame. When declared a \qasmline{frame} is tied to a \qasmline{port}. Frames may be defined both within the program scope --- \qasmline{frame frd0 = newframe(d0, 5e9, 0.0);} --- or as an external linkage to plugin to existing calibration frameworks --- \qasmline{extern frd0;}. A frame contains two attributes, \qasmline{frequency} and \qasmline{phase} of types \qasmline{float} and \qasmline{angle} respectively. These may be set and modified through dot notation. For example, \qasmline{frd0.frequency = 4.9e9; frd0.phase += π;} sets the frequency for a frame and increments the phase of the frame. These updates will occur at the current time of the frame. Multiple frames may be defined for each port effectively allowing multiplexing over a port's IO for multiple carrier signals. Consequently, each frame maintains its pointer to an abstract internal clock which is incremented through operations such as \qasmline{play} and \qasmline{barrier}. 

The following code sample demonstrates how to define an implementation of an \qasmline{rz} gate for the physical qubit \qasmline{$0}.

\begin{qasmblock}
defcal rz(angle θ) $0 {
    driveframe0.phase -= θ;
}
\end{qasmblock}
\noindent
In this example, the gate is realized by updating the frame tracking the qubits. All subsequent pulses that we apply to the frame will have an updated phase, allowing the implementation of the virtual Z-gate.

The manipulation of a \qasmline{port} is channeled through the \qasmline{frame} which provides a \textit{view} on top of the \qasmline{port}. The primary stimulus applied to a \qasmline{frame} is the \qasmline{waveform} type. These may be defined either as an array of complex samples which define the points for the waveform envelope --- \qasmline{waveform arb_waveform = [1+0im, 0+1im, 0.5+0.5im]}. Or as an
\qasmline{extern} declared waveform template representing a waveform which will later be materialized into a list of complex samples, either by the compiler or the hardware --- \qasmline{extern gaussian(complex[float] amp, duration d, duration sigma) -> waveform;}. Waveforms are played on a frame through the \qasmline{play} instruction 

\begin{qasmblock}
defcal sx $0 {
    // emit a pulse with a carrier signal, time and
    // port defined by frd0.
    play(gaussian(0.1, 160dt, 40dt), frd0);
}
\end{qasmblock}

The \qasmline{waveform} is therefore emitted with a carrier defined by the frequency and phase of \qasmline{frd0} at the current time of \qasmline{frd0} on port \qasmline{d0}. The frame is blocked during execution of the \qasmline{play} operation and its time is incremented by \qasmline{160dt} after execution completes.

It is also possible to calibrate the \qasmline{measure} and \qasmline{reset} operations. Due to the varied nature of readout protocols in quantum hardware we rely on vendors to define the appropriate external subroutines to access the readout chain. For example, an implementation of \qasmline{measure} is shown below.
\begin{qasmblock}
cal {
    // Capture definition available in the hardware
    extern capture(duration dur, frame capture_frame) -> bit;
  
    extern port m0;
    extern port cap0;
  
    frame frm0 = newframe(m0, 5e9, 0.);
    frame frcap0 = newframe(cap0, 5e9, 0.);
}

defcal measure $0 -> bit {
    // measurement stimulus envelope
    waveform meas_wf = gaussian_square(0.1, 16000dt, 262dt, 13952dt);

    // play the stimulus
    play(meas_wf, frm0);
    // delay the capture frame to account for signal latencies
    delay[320dt] frcap0;
    // capture transmitted data after interaction with measurement resonator
    return capture(16000dt, frcap0);
}
\end{qasmblock}

OpenPulse's timing model is \textit{relative}. Within a program, there is no explicit reference to a global clock, but instead, only relative references to the starting time of a \qasmline{defcal/cal} or the current relative time of other frames through the \qasmline{barrier} instruction. This enables OpenPulse operations to be scheduled in a \textit{position-independent} manner. Each \qasmline{defcal} or \qasmline{cal} block defines a relative zero-time which is defined by when the OpenQASM scheduling layer schedules the execution of the block. When a \qasmline{frame} is declared it is assigned an absolute time that is equal to the relative zero time for the enclosing block. Every \qasmline{frame} used within a \qasmline{cal} or \qasmline{defcal} starts and ends with an implicit barrier across all used frames, establishing a unified start and end time for all frames within the block. The example below demonstrates the timing behaviour of OpenPulse with the implementation of an echoed cross-resonance gate \cite{sheldon2016procedure}.

\begin{qasmblock}
cal {
    extern port d0;
    extern frame frd0;
    extern frame frd1;
    extern frame frd2;
}

defcal ecr $0 $1 {
    // define and update frame at t=0
    frame cr1_0 = newframe(d0, frd1.frequency, frd1.phase);
    // synchronize time of frames at t1=max_time(frd0, frd1, frd2)
    barrier frd0, frd1, frd2, cr1_0;
    // play a waveform on frame cr1_0 incrementing time of cr1_0 by dur
    play(gaussian_square(dur, -amp, sigma, square_width), cr1_0);
    // synchronize frames before next call at t2=t1+dur
    barrier frd0, frd1, frd2, cr1_0;
    // call another defcal
    x $0;
    // synchronize frames before next call at t3=t2+duration(x)
    barrier frd0, frd1, frd2, cr1_0;
    play(gaussian_square(dur, -amp, sigma, square_width), cr1_0);
    // synchronize frames before leaving defcal at t4=t3+dur
    barrier frd0, frd1, frd2, cr1_0;
}
\end{qasmblock}

\noindent
The example above demonstrates several additional features of the OpenPulse language. The notion and semantics of the \qasmline{barrier frd0, frd1, frd2, cr1_0;} statement is borrowed from the circuit layer to allow the OpenPulse grammar extension to synchronize the time of all frames to a unique instant. A \qasmline{defcal} may also call another \qasmline{defcal} to construct a more sophisticated calibration as shown by the call of \qasmline{x $0;}. Furthermore, the use of the external frame \qasmline{frd2} enables the calibration grammar to inform the circuit scheduler that qubit \qasmline{$2} is indirectly used by the gate. The scheduler is therefore made aware that the qubits are indirectly coupled. This allows it to avoid scheduling them simultaneously, such as the case where many qubits are coupled through a resonator bus.

\subsection{Solving the stretch problem}
\label{sec:stretch-problem}

We now describe how the compiler determines the values of \qasmline{stretch} values, in contexts involving operations, with other \qasmline{duration} values, and \qasmline{box} statements.
We treat \qasmline{duration} values to be variables
in a linear system of inequalities that the compiler must solve, subject to certain constraints.
We call this problem the ``stretch problem''.
If we attempt to execute an OpenQASM circuit, a stretch solving algorithm must be applied to compute explicit durations for all delays.

Our goal is that any OpenQASM circuit with mapped qubits and \qasmline{defcal}s gives rise to a set of stretch problems that, if bounded and feasible, provides explicit timing for all operations within every basic block.
The stretch problem is to determine definite values for all \qasmline{stretch} values in a given scope --- subject to constraints on gate ordering, such as those imposed by \qasmline{delay} and \qasmline{barrier} statements, and the range of optimization techniques --- so that the operations performed on each qubit can be scheduled appropriately, and in particular take the same amount of time.
The stretches get resolved to explicit delays in a single late-stage pass that solves the stretch problem, thereby specifying all delays and a complete schedule for the operations. We define the stretch problem in such a way that it is bounded, and has a unique optimum if it is feasible. If the problem is infeasible for a given target, a compiler error is issued, providing some insight about the unsatisfied constraints.
In the case of a circuit without any stretchable delays, the stretch problem may be unsolvable in all but the most trivial cases.

The stretch problem is a lexicographic multi-objective linear programming problem~\cite{CPS18}, as follows: given a matrix $A$ and vectors $c_j$ and $b$, lexicographically minimize
$c_1^\top x$, $c_2^\top x$, ..., $c_m^\top x$ subject to $Ax\leq b$ and $x\geq 0$. This linear programming
problem can be solved in polynomial time and efficiently in practice. For example, one may first solve ``minimize $c_1^\top x$ subject to $Ax\leq b$ and $x\geq 0$'' to obtain the cost $\beta_1$, then solve ``minimize $c_2^\top x$ subject to $Ax\leq b$, $c_1^\top x=\beta_1$, and $x\geq 0$'' to obtain the cost $\beta_2$, and for forth until all objectives are met. 
This is described in \cite{CPS18} and the references therein;  multiobjective optimization is implemented in software such as CPLEX~\cite{cplex2009v12}.

The stretch problem is formulated independently for each basic block or box $\mathrm{B}$. 
The total duration of the block is mapped to some variable $\mathrm T = x_0$, which we wish to minimize.
The block $\mathrm{B}$  consists of instructions applied to qubits $\mathrm{Q}$.
An instruction is any task with a non-negative duration 
that involves a subset of qubits; the duration of each instruction is either known or unknown, and may be associated with variables $x_1$, $x_2$, \emph{etc.} 
To clarify the presentation of the stretch problem, imagine that each qubit $q \in \mathrm{Q}$ is associated with a duration value $D_q$\,.
The value of $D_q$ is determined by the durations of operations and \qasmline{delays} on $q$, as well as constraints on when multi-qubit blocks can be performed (depending on the scheduling of operations on the that must come before, on each of the qubits involved).
This defines $D_q$ a function of the non-negative stretch variables $x_i$. 
We then impose the constraint that $x_0 = D_q$ for all $q \in \mathrm{Q}$\,.
A feasible solution to this assigns durations to each \qasmline{stretch} variable, ensuring that for each qubit $q \in \mathrm{Q}$, the operations on $q$ occur at some time $t\in [0, \mathrm{T}]$.
If we lexicographically minimize the variables $x_i$, we single out a particular solution where, among other things, $\mathrm{T} = x_0$ is minimized.

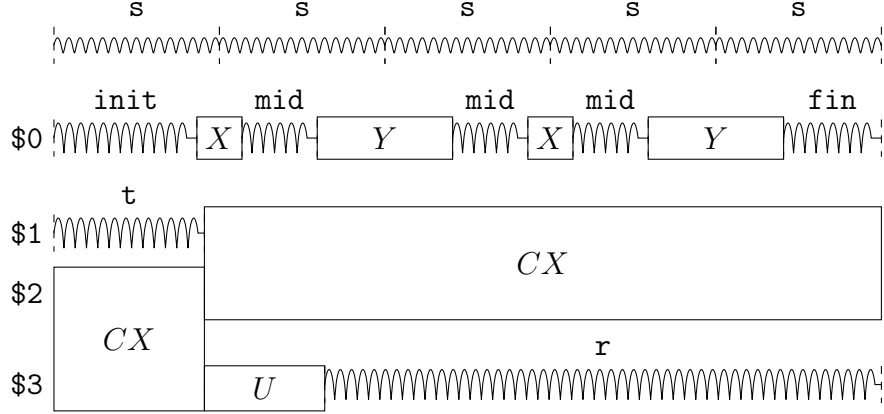
\begin{figure*}[h]
    \centering
    \begin{tikzpicture}
        \def \xlen {3mm}
        \def \ylen {9mm}
        \def \eqlen {11mm}
        \def \cxlen {10mm}
        \def \ulen {8mm}
        \newlength{\startlen}
        \newlength{\midlen}
        \newlength{\termlen}
        \newlength{\cxxlen}
        \newlength{\tlen}
        \newlength{\rlen}
        \setlength{\startlen}{\dimexpr(\eqlen-\xlen/2)\relax}
        \setlength{\midlen}{\dimexpr(\eqlen-\xlen/2-\ylen/2)\relax}
        \setlength{\termlen}{\dimexpr(\eqlen-\ylen/2)\relax}
        \setlength{\cxxlen}{\dimexpr(\eqlen*5-\cxlen)\relax}
        \setlength{\tlen}{\dimexpr(\cxlen)\relax}
        \setlength{\rlen}{\dimexpr(\eqlen*5-\cxlen-\ulen)\relax}
      \begin{yquant*}[operator/separation=0mm, register/separation=2mm]
        nobit p;
        qubit {\texttt{\$\idx}} q[4];
          %hspace {5mm} p;
          ["\texttt s", name=eq0, x radius=\eqlen, draw=none] box {} p;
          ["\texttt s", name=eq1, x radius=\eqlen, draw=none] box {} p;
          ["\texttt s", name=eq2, x radius=\eqlen, draw=none] box {} p;
          ["\texttt s", name=eq3, x radius=\eqlen, draw=none] box {} p;
          ["\texttt s", name=eq4, x radius=\eqlen, draw=none] box {} p;
          ["\texttt{init}", name=ss, x radius=\startlen, draw=none] box {} q[0];
          [x radius=\xlen] box {$X$} q[0];
          ["\texttt{mid}", name=ms0, x radius=\midlen, draw=none] box {} q[0];
          [x radius=\ylen] box {$Y$} q[0];
          ["\texttt{mid}", name=ms1, x radius=\midlen, draw=none] box {} q[0];
          [x radius=\xlen] box {$X$} q[0];
          ["\texttt{mid}", name=ms2, x radius=\midlen, draw=none] box {} q[0];
          [x radius=\ylen] box {$Y$} q[0];
          ["\texttt{fin}", name=es, x radius=\termlen, draw=none] box {} q[0];
          ["\texttt t", name=t, x radius=\tlen, draw=none] box {} q[1];
          [x radius=\cxlen, y radius=7mm] box {$CX$} (q[2], q[3]);
          [x radius=\cxxlen, y radius=7mm] box {$CX$} (q[1], q[2]);
          [x radius=\ulen, y radius=3mm, yshift=-.5mm] box {$U$} (q[3]);
          ["\texttt r", name=r, x radius=\rlen, draw=none] box {} q[3];
      \end{yquant*}
      \spring{eq0}{1mm}
      \spring{eq1}{1mm}
      \spring{eq2}{1mm}
      \spring{eq3}{1mm}
      \spring{eq4}{1mm}
      \spring{ss}{2mm}
      \spring{ms0}{2mm}
      \spring{ms1}{2mm}
      \spring{ms2}{2mm}
      \spring{es}{2mm}
      \spring{t}{2mm}
      \spring{r}{2mm}
    \end{tikzpicture}
    \caption{Dynamical decoupling of a spectator qubit using finite-duration pulses.
    This design intent can be expressed by defining a single stretch variable $s$ that corresponds to the
    distance between equidistant gate centers. The other durations which correspond to actual circuit delays
    are derived by simple arithmetic on durations. Given a target system with calibrated X and Y gates,
    the solution to the stretch problem can be found.}
    \label{fig:dd}
\end{figure*}

For example, the code below inserts a dynamical decoupling sequence where the \emph{centers} of pulses are equidistant from each other.
The circuit is depicted in Figure~\ref{fig:dd}.
We specify correct durations for the delays by using backtracking operations to properly take into account the finite duration of each gate.

\qasmfilewithlines{examples/stretch-dd.qasm}

\noindent
The \qasmline{stretch} declarations impose the constraints:
$\texttt{s},\, \texttt{t},\, \texttt{r},\, \texttt{xlen},\, \texttt{ylen} \geq 0$,
$\texttt{s}-\tfrac{1}{2}\,\texttt{xlen} \geq 0$,
$\texttt{s} - \tfrac{1}{2}\,\texttt{xlen} - \tfrac{1}{2}\,\texttt{ylen} \geq 0$,
and
$\texttt{s} - \tfrac{1}{2}\,\texttt{ylen} \geq 0$.
Some additional constraints $\texttt{xlen} \ge A$ and $\texttt{ylen} \ge B$ are imposed by the target platform, where $A$ and $B$ are the minimum amount of time possible  to realize the operations \,\qasmline{x $0}\, and \,\qasmline{y $0}\, respectively.
Let $x_{16}$, $x_{17}$, and $x_{18}$ denote the (fixed and not explicitly declared) durations of the operations on lines 16, 17, and 18.
Then we obtain the following durations of operations on qubits:
\begin{equation}
    \begin{aligned}
        D_{\texttt{q[0]}} \,&=\, 5\,\texttt{s}\,,
    \\
        D_{\texttt{q[1]}} \,&=\, \texttt{t} + x_{17}\,,
    \end{aligned}
    \qquad\qquad\qquad
    \begin{aligned}
        D_{\texttt{q[2]}} \,&=\, x_{16} + x_{17}\,,
    \\
        D_{\texttt{q[3]}} \,&=\, x_{16} + x_{18} + \texttt{r}\,.
    \end{aligned}
\end{equation}
By setting $D_{\texttt{q[i]}} = \mathrm{T}$ for all $\texttt{i}$, we obtain additional relationships and constraints on the durations above, specifically:
\begin{equation}
    \begin{aligned}
        \textrm{T} \,&=\, x_{16} + x_{17},
    &\qquad\quad
        \texttt{s} \,&=\, \tfrac{1}{5}x_{16} + \tfrac{1}{5}x_{17},
    &\qquad\quad
        \texttt{t} \,&=\, x_{16},
    &\qquad\quad 
        \texttt{r} \,&=\, x_{17} - x_{18}.
    \end{aligned}
\end{equation}
From the constraints on $\texttt{s}$ and $\texttt{r}$, this is feasible if $x_{16} + x_{17} \ge \tfrac{5}{2}A + \tfrac{5}{2}B$ and $x_{17} \ge x_{18}$\,.
The solution which is obtained by the stretch problem in this case actually determines fixed values for $\textrm{T}$, $\texttt{s}$, $\texttt{t}$, and $\texttt{r}$, and sets $\texttt{xlen}$ to its minimum value $A$, and $\texttt{ylen}$ to its minimum value $B$. 

Note that the order in which \qasmline{stretch} variables are declared determine the order in which they are minimized.
This can have a significant effect on scheduling.
For a second example, consider the following code sample:
\qasmfilewithlines{examples/stretch-order.qasm}
\noindent 
Let $x_4$ and $x_5$ stand for the (fixed and not explicitly declared) durations of the operations on lines $4$ and $5$.
We have
\begin{equation}
\begin{aligned}
    \mathrm{lexmin}\ \texttt{a},\, \texttt{b},\, \texttt{c},\, \texttt{d}\ & \ \mathrm{s.t.} \\
    \texttt{a}, \texttt{b}, \texttt{c}, \texttt{d} & \geq 0 & \\
    \texttt{a} + x_4 + \texttt{c} & = \texttt{b} + x_5 + \texttt{d}.
\end{aligned}
\end{equation}
From this, optimising first for $\texttt{a}$, then $\texttt{b}$, and so forth, we would obtain $\texttt{a} = \texttt{b} = 0$, and either $\texttt{c} = x_5 - x_4$ and $\texttt{d} = 0$ if $x_4 \leq x_5$, or $\texttt{c} = 0$ and $\texttt{d} = x_4 - x_5$ otherwise.
This yields a left-aligned schedule.
If we had instead declared \qasmline{stretch d, c, b, a;} on line~1, we would have optimized the values in the opposite order (minimising first $\texttt{d}$, then $\texttt{c}$, \emph{etc.}) to obtain a right-aligned schedule.

\subsection{Multi-level representation}
\label{sec:multilevel}

In previous sections we introduced the language features of OpenQASM, which
can range from relatively high-level constructs such as multi-controlled gates to low-level
timing and microcoded pulses. As such, OpenQASM is designed to be a multi-level 
intermediate representation, where the focus shifts from target-agnostic computation to
concrete implementation as more hardware specificity is introduced.

An OpenQASM circuit can also mix different abstraction levels by introducing constraints where
needed, but allowing the compiler to make decisions where there are no constraints.
Examples of circuit constraints are tying a virtual qubit to a particular physical qubit,
or left-aligning some parallel gates that have different durations.

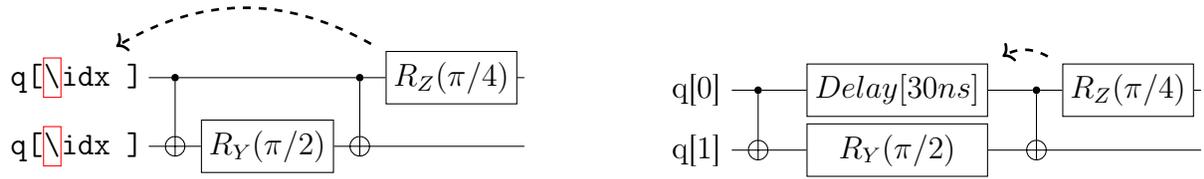
\begin{figure*}
    \centering
    \begin{subfigure}[b]{.2\textwidth}
        \centering
        \begin{tikzpicture}
          \begin{yquant}
              [name=ypos]
              qubit {\qasmline{\reg[\idx]}\!} q[2];
              cnot q[1] | q[0];
              box {$R_Y(\pi/2)$} q[1];
              cnot q[1] | q[0];
              [name=rz] box {$R_Z(\pi/4)$} q[0];
          \end{yquant}
          \draw [line width=1pt, ->, dashed]
          ([yshift=1mm, xshift=-2mm] rz-0.north west)
          to [bend right]
          (ypos-0);
        \end{tikzpicture}
    \end{subfigure}
    \hspace*{\fill}
    \begin{subfigure}[b]{.5\textwidth}
        \centering
        \begin{tikzpicture}
          \begin{yquant}
              qubit q[2];
              cnot q[1] | q[0];
              hspace {3mm} q;
              [name=del, x radius=12mm] box {$Delay [30ns]$} q[0];
              [x radius=12mm] box {$R_Y(\pi/2)$} q[1];
              hspace {3mm} q;
              cnot q[1] | q[0];
              [name=rz] box {$R_Z(\pi/4)$} q[0];
          \end{yquant}
            \draw [line width=1pt, ->, dashed] ([yshift=1mm, xshift=-2mm]rz-0.north west) to [bend right] ([yshift=1mm, xshift=2mm] del-0.north east);
        \end{tikzpicture}
    \end{subfigure}
    \caption{Implicit vs. explicit delay. (a) An implicit delay exists on $q[0]$, but it is not part of the circuit description. Thus this circuit does not care about timing and the $RZ$ gate is free to commute on the top wire. (b) An explicit delay is part of the circuit description. The timing is consistent and can be resolved if and only if this delay is exactly the same duration as $RY$ on $q[1]$. The delay is like a barrier in that it prevents commutation on that wire. However $RZ$ can still commute before the $CNOT$ if it has duration $0$.}
    \label{fig:delaycommute}
\end{figure*}

We illustrate this point by considering timing constraints in an OpenQASM circuit.
When a \qasmline{delay} instruction is used, even though it implements the identity channel in
the ideal case, it is understood to provide explicit timing. Therefore an explicit \qasmline{delay}
instruction will prevent commutation of gates that would otherwise commute. For example in 
Figure~\ref{fig:delaycommute}a, there will be an implicit delay between the `\qasmline{cx}' gates 
on qubit 0. However, the `\qasmline{rz}' gate is still free to commute on that qubit, because the 
delay is implicit. Once the delay becomes explicit (perhaps at lower stages of compilation), gate 
commutation is prohibited (Figure~\ref{fig:delaycommute}b).

Furthermore, even though stretch is used to specify constraints for the solver,
its use in the circuit is like any other instruction. This has the benefit that design intents can
be specified on a high-level circuit and the intent will be carried with the circuit, until resolved.
The following example for simultaneous randomized benchmarking is an instance of this,
where alignment is specified on general unitaries of the Clifford group, which
may be decomposed in a variety of ways. In addition, the stretchiness can be included
as part of the definition of a gate, in which case whenever that gate is expanded those 
stretch constraints will be automatically inserted.

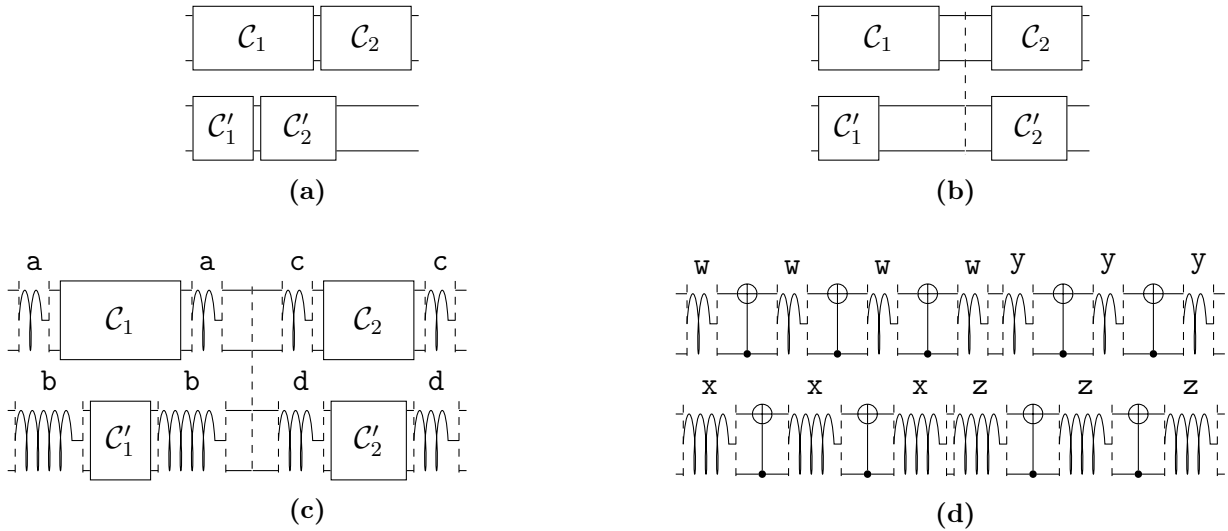
\begin{figure*}
    \centering
    \hspace{-5mm}
    \begin{subfigure}{.5\textwidth}
    \centering
\scalebox{1}{
\begin{tikzpicture}
  \begin{yquant}
        [register/minimum height=5mm]
      qubit {} q[4];
      [x radius=8mm] box {$\mathcal{C}_1$} (q[0], q[1]);
      [x radius=4mm] box {$\mathcal{C}_1^\prime$} (q[2], q[3]);
      [x radius=6mm] box {$\mathcal{C}_2$} (q[0], q[1]);
      [x radius=5mm] box {$\mathcal{C}_2^\prime$} (q[2], q[3]);
  \end{yquant}
\end{tikzpicture}
}
        \label{fig:none}
        \caption{}
    \end{subfigure}
    \hfill
    \begin{subfigure}{.5\textwidth}
    \centering
\scalebox{1}{
\begin{tikzpicture}
  \begin{yquant}
        [register/minimum height=5mm]
      qubit {} q[4];
      [x radius=8mm] box {$\mathcal{C}_1$} (q[0], q[1]);
      [x radius=4mm] box {$\mathcal{C}_1^\prime$} (q[2], q[3]);
      barrier (q[0], q[1], q[2], q[3]);
      [x radius=6mm] box {$\mathcal{C}_2$} (q[0], q[1]);
      [x radius=5mm] box {$\mathcal{C}_2^\prime$} (q[2], q[3]);
  \end{yquant}
\end{tikzpicture}
}
        \label{fig:barrier}
        \caption{}
    \end{subfigure}

    \bigskip
    \hspace{-5mm}
    \begin{subfigure}{.5\textwidth}
\scalebox{1}{
\begin{tikzpicture}
  \begin{yquant}
      [register/minimum height=7mm]
      qubit {} q[4];
      ["\texttt a", name=boxa, x radius=2mm, draw=none] box {} (q[0], q[1]);
      [x radius=8mm] box {$\mathcal{C}_1$} (q[0], q[1]);
      ["\texttt a", name=boxb, x radius=2mm, draw=none] box {} (q[0], q[1]);
      ["\texttt b", name=boxc, x radius=4.5mm, draw=none] box {} (q[2], q[3]);
      [x radius=4mm] box {$\mathcal{C}_1^\prime$} (q[2], q[3]);
      ["\texttt b", name=boxd, x radius=4.5mm, draw=none] box {} (q[2], q[3]);
      barrier (q[0], q[1], q[2], q[3]);
      ["\texttt c", name=boxe, x radius=2mm, draw=none] box {} (q[0], q[1]);
      [x radius=6mm] box {$\mathcal{C}_2$} (q[0], q[1]);
      ["\texttt c", name=boxf, x radius=2mm, draw=none] box {} (q[0], q[1]);
      ["\texttt d", name=boxg, x radius=3mm, draw=none] box {} (q[2], q[3]);
      [x radius=5mm] box {$\mathcal{C}_2^\prime$} (q[2], q[3]);
      ["\texttt d", name=boxh, x radius=3mm, draw=none] box {} (q[2], q[3]);
  \end{yquant}
  \spring{boxa}{4mm};
  \spring{boxb}{4mm};
  \spring{boxc}{4mm};
  \spring{boxd}{4mm};
  \spring{boxe}{4mm};
  \spring{boxf}{4mm};
  \spring{boxg}{4mm};
  \spring{boxh}{4mm};
\end{tikzpicture}
}
        \label{fig:mid}
        \caption{}
    \end{subfigure}
    \hfill
    \begin{subfigure}{.5\textwidth}
    \centering
\scalebox{1}{
\begin{tikzpicture}
  \begin{yquant}
      [register/minimum height=7mm]
      qubit {} q[4];
      ["\texttt w", name=box1a, x radius=2mm, draw=none] box {} (q[0], q[1]);
      cnot q[0] | q[1];
      ["\texttt w", name=box1b, x radius=2mm, draw=none] box {} (q[0], q[1]);
      cnot q[0] | q[1];
      ["\texttt w", name=box1c, x radius=2mm, draw=none] box {} (q[0], q[1]);
      cnot q[0] | q[1];
      ["\texttt w", name=box1d, x radius=2mm, draw=none] box {} (q[0], q[1]);
      ["\texttt x", name=box2a, x radius=3.5mm, draw=none] box {} (q[2], q[3]);
      cnot q[2] | q[3];
      ["\texttt x", name=box2b, x radius=3.5mm, draw=none] box {} (q[2], q[3]);
      cnot q[2] | q[3];
      ["\texttt x", name=box2c, x radius=3.5mm, draw=none] box {} (q[2], q[3]);
      ["\texttt y", name=box3a, x radius=2mm, draw=none] box {} (q[0], q[1]);
      cnot q[0] | q[1];
      ["\texttt y", name=box3b, x radius=2mm, draw=none] box {} (q[0], q[1]);
      cnot q[0] | q[1];
      ["\texttt y", name=box3c, x radius=2mm, draw=none] box {} (q[0], q[1]);
      ["\texttt z", name=box4a, x radius=3.5mm, draw=none] box {} (q[2], q[3]);
      cnot q[2] | q[3];
      ["\texttt z", name=box4b, x radius=3.5mm, draw=none] box {} (q[2], q[3]);
      cnot q[2] | q[3];
      ["\texttt z", name=box4c, x radius=3.5mm, draw=none] box {} (q[2], q[3]);
  \end{yquant}
  \spring{box1a}{4mm};
  \spring{box1b}{4mm};
  \spring{box1c}{4mm};
  \spring{box1d}{4mm};
  \spring{box2a}{4mm};
  \spring{box2b}{4mm};
  \spring{box2c}{4mm};
  \spring{box3a}{4mm};
  \spring{box3b}{4mm};
  \spring{box3c}{4mm};
  \spring{box4a}{4mm};
  \spring{box4b}{4mm};
  \spring{box4c}{4mm};
\end{tikzpicture}
}
        \label{fig:cnots}
        \caption{}
    \end{subfigure}
    
    \caption{Simultaneous 2-qubit randomized benchmarking using stretchy delays to align gates. (a) Independent sequences of Clifford operations are applied to the top 2 and bottom 2 qubits, but varying decompositions and gate speeds cause misalignment. (b) A barrier can force coarse synchronization points. (c) Stretchy delays inserted before and after every Clifford can ensure better simultaneity. These delays will be carried as part of the circuit, no matter how the Cliffords are decomposed. (d) The decomposition of Cliffords themselves can have stretchy delays, ensuring even finer alignment of gates. The program can be written in a way that these delays override the previous ones in-between Cliffords.
    \label{fig:srb}}
\end{figure*}

\section{Compilation phases by example}

In this section we present a full example using an OpenQASM circuit for an iterative phase estimation algorithm~\cite{iqpe}. We show how the circuit may get transformed during multiple phases of compilation, and how OpenQASM can be used as the intermediate representation at each phase. This simple example also serves to highlight many of the features of the language: classical control flow, gate modifiers, virtual and physical qubits, timing and stretches, and pulse-defined calibrations.

OpenQASM itself is agnostic to any particular toolchain, so the compilation phases and transformations in this section are merely intended as a representative example of typical steps in circuit compilation.

Iterative phase estimation is an algorithm for calculating the eigenvalue (phase) of a unitary up to some number of bits of precision. In contrast to the textbook phase estimation, it uses fewer qubits (only one control qubit), but at the cost of multiple measurements. Each measurement adds one bit to the estimated phase. Importantly, all iterations must happen in real time during the coherence interval of the qubits, and the result of each measurement feeds forward to the following iteration to influence the angles of rotation.

\vspace*{-1ex}
\paragraph{Initial circuit.} The initial circuit may have been produced by high level software tools and languages and may have already passed through stages of high level transformations.

\qasmfile{examples/example-1.qasm}
\noindent\makebox[\textwidth]{
\scalebox{1}{
\begin{tikzpicture}
      \def \niter {\ensuremath{n}};
   \begin{yquant*}
      qubit {\qasmline{\reg}\!} q;
      qubit {\qasmline{\reg}\!} r;
      cbit {\qasmline{c=0}} c;
      ["north east:\niter" {font=\protect\small, inner sep=0pt}]
      slash c;
      [draw=none] box {$\ket{0}$} q;
      [draw=none] box {$\ket{0}$} r;
      h r;
      [this subcircuit box style={draw=none, inner sep=0pt, "$0\,..\,\niter{-}1$"}]
      repstart (q, r);
      [draw=none] box {$\ket{0}$} q;
      h q;
      box {$P(\theta)^{2^k}$} r | q;
      box {$P(c)^{-1}$} q;
      h q;
      measure q;
      %not c[0] | q;
      [operators/every barrier/.append style={-{Latex[length=2.5mm]}, double, double distance=.4mm, line width=.15mm, solid, shorten <= 2.5mm, shorten >= 2.5mm}, operator/separation=1pt]
    barrier (q, r, c[0]);
      settype {qubit} q;
      [this subcircuit box style={draw=none, align=center, inner sep=0pt, "$k++$"}]
      repend (q, r);
      box {$c \ll= 1$} c;
   \end{yquant*}
\end{tikzpicture}
}
}

\subsection{Target-independent compilation phase}

The target-independent phase applies transformations that do not use information about any particular target system. The goal of these transformations is quantum circuit synthesis and optimization.

\vspace*{-1ex}
\paragraph{Constant propagation and folding.} This transformation substitutes the values of constants as they occur in expressions throughout the circuit. We repeatedly apply this transformation as constants are exposed.

\vspace*{-1ex}
\paragraph{Gate modifier evaluation and synthesis.} The control, power, and inverse gate modifiers are replaced by gate sequences, and those are simplified when possible based on the structure of the circuit. In this case, the inverse and power modifiers can be simplified by modifying the angle arguments of the phase and controlled-phase gates. For this example, we replace the control modifier and \qasmline{phase} gate by a \qasmline{cphase} gate from the standard library to emphasize the transformation. For further emphasis, we also change the circuit notation, replacing a controlled unitary gate (raised to a power) by a two-qubit gate with two connected control dots and an angle label.

\qasmfile{examples/example-3.qasm}

\noindent\makebox[\textwidth]{
\scalebox{.9}{
\begin{tikzpicture}
      \def \niter {3};
   \begin{yquant*}
      qubit {\qasmline{\reg}\!} q;
      qubit {\qasmline{\reg}\!} r;
      cbit {\qasmline{c=0}} c;
      ["north east:\niter" {font=\protect\small, inner sep=0pt}]
      slash c;
      [draw=none] box {$\ket{0}$} q;
      [draw=none] box {$\ket{0}$} r;
      h r;
      [this subcircuit box style={draw=none, inner sep=0pt, "\niter $\times$"}]
      repstart (q, r);
      [draw=none] box {$\ket{0}$} q;
      h q;
      phase {$2^k \cdot \frac{3\pi}{8}$} r | q;
      phase {$-c$} q;
      h q;
      measure q;
      %not c[0] | q;
      [operators/every barrier/.append style={-{Latex[length=2.5mm]}, double, double distance=.4mm, line width=.15mm, solid, shorten <= 2.5mm, shorten >= 2.5mm}, operator/separation=1pt]
    barrier (q, r, c[0]);
      settype {qubit} q;
      [this subcircuit box style={draw=none, align=center, inner sep=0pt, "$k++$"}]
      repend (q, r);
      box {$c \ll= 1$} c;
   \end{yquant*}
\end{tikzpicture}
}}

\vspace*{-1ex}
\paragraph{Loop unrolling.} The trip count of this loop is constant and statically known. Therefore the loop can be unrolled. This is not necessary as the machine executing an OpenQASM circuit is assumed to have control flow capabilities. However, doing so in the compiler could expose certain optimization opportunities. For example, the following pass could remove the reset operation in the first iteration, because it is redundant with the initial circuit reset.

\qasmfile{examples/example-4.qasm}
\noindent\makebox[\textwidth]{
\scalebox{.7}{
\begin{tikzpicture}
   \begin{yquant*}
      qubit {\qasmline{\reg}\!} q;
      qubit {\qasmline{\reg}\!} r;
      cbit {\qasmline{\reg[\idx]} = 0\!} c[3];
      [draw=none] box {$\ket{0}$} q;
      [draw=none] box {$\ket{0}$} r;
      h r;
      h q;
      phase {$2^k\cdot\frac{3\pi}{8}$} r | q;
      phase {$-c$} q;
      h q;
      measure q;
      %not c[0] | q;
      [operators/every barrier/.append style={-{Latex[length=2.5mm]}, double, double distance=.4mm, line width=.15mm, solid, shorten <= 2.5mm, shorten >= 2.5mm}, operator/separation=1pt]
    barrier (q, r, c[0]);
      box {$c \ll= 1$} (c[0], c[1], c[2]);
      settype {qubit} q;
      [draw=none] box {$\ket{0}$} q;
      h q;
      phase {$2^k\cdot\frac{3\pi}{8}$} r | q;
      phase {$-c$} q;
      h q;
      measure q;
      %not c[0] | q;
      [operators/every barrier/.append style={-{Latex[length=2.5mm]}, double, double distance=.4mm, line width=.15mm, solid, shorten <= 2.5mm, shorten >= 2.5mm}, operator/separation=1pt]
    barrier (q, r, c[0]);
      box {$c \ll= 1$} (c[0], c[1], c[2]);
      settype {qubit} q;
      [draw=none] box {$\ket{0}$} q;
      h q;
      phase {$2^k\cdot\frac{3\pi}{8}$} r | q;
      phase {$-c$} q;
      h q;
      measure q;
      %not c[0] | q;
      [operators/every barrier/.append style={-{Latex[length=2.5mm]}, double, double distance=.4mm, line width=.15mm, solid, shorten <= 2.5mm, shorten >= 2.5mm}, operator/separation=1pt]
    barrier (q, r, c[0]);
      box {$c \ll= 1$} (c[0], c[1], c[2]);
      settype {qubit} q;
   \end{yquant*}
\end{tikzpicture}
}}

\vspace*{-1ex}
\paragraph{Constant propagation and folding (repeat).} We repeat this step to evaluate and substitute the value of the \qasmline{power} variable.

\vspace*{-1ex}
\paragraph{Gate simplification.} Gate simplification rules can be applied to reduce the gate count. In this example, rotations with angle zero are identity gates and can be removed.

\qasmfile{examples/example-5.qasm}
\noindent\makebox[\textwidth]{
\scalebox{.8}{
\begin{tikzpicture}
   \begin{yquant*}
      qubit {\qasmline{\reg}\!} q;
      qubit {\qasmline{\reg}\!} r;
      cbit {\qasmline{\reg[\idx]} = 0\!} c[3];
      [draw=none] box {$\ket{0}$} q;
      [draw=none] box {$\ket{0}$} r;
      h r;
      h q;
      phase {$\frac{3\pi}{8}$} r | q;
      h q;
      measure q;
      %not c[0] | q;
      [operators/every barrier/.append style={-{Latex[length=2.5mm]}, double, double distance=.4mm, line width=.15mm, solid, shorten <= 2.5mm, shorten >= 2.5mm}, operator/separation=1pt]
    barrier (q, r, c[0]);
      box {$c \ll= 1$} (c[0], c[1], c[2]);
      settype {qubit} q;
      [draw=none] box {$\ket{0}$} q;
      h q;
      phase {$\frac{6\pi}{8}$} r | q;
      phase {$-c$} q;
      h q;
      measure q;
      %not c[0] | q;
      [operators/every barrier/.append style={-{Latex[length=2.5mm]}, double, double distance=.4mm, line width=.15mm, solid, shorten <= 2.5mm, shorten >= 2.5mm}, operator/separation=1pt]
    barrier (q, r, c[0]);
      box {$c \ll= 1$} (c[0], c[1], c[2]);
      settype {qubit} q;
      [draw=none] box {$\ket{0}$} q;
      h q;
      phase {$\frac{12\pi}{8}$} r | q;
      phase {$-c$} q;
      h q;
      measure q;
      %not c[0] | q;
      [operators/every barrier/.append style={-{Latex[length=2.5mm]}, double, double distance=.4mm, line width=.15mm, solid, shorten <= 2.5mm, shorten >= 2.5mm}, operator/separation=1pt]
    barrier (q, r, c[0]);
      box {$c \ll= 1$} (c[0], c[1], c[2]);
      settype {qubit} q;
   \end{yquant*}
\end{tikzpicture}
}}

\subsection{Target-dependent compilation phase}

The target-dependent phase applies transformations that are generically necessary to lower a target-independent OpenQASM 3 circuit to an executable. Each transformation uses information that is specific to a class of target machines, rather than one specific target, and the class of potential targets is refined with each transformation.

\vspace*{-1ex}
\paragraph{Basis translation.} As we enter the next phase of compilation, we need to use information about the target system. The basis translation step rewrites quantum gates in terms of a set of gates available on the target. In this example, we assume a quantum computer with a calibrated set of basis gates including \qasmline{phase}, \qasmline{h}, and \qasmline{cx}.

\qasmfile{examples/example-6.qasm}
\noindent\makebox[\textwidth]{
\scalebox{.7}{
\begin{tikzpicture}
   \begin{yquant*}[operator/separation=0mm]
      qubit {\qasmline{\reg}\!} q;
      qubit {\qasmline{\reg}\!} r;
      cbit {\qasmline{\reg[\idx]} = 0\!} c[3];
      [draw=none] box {$\ket{0}$} q;
      [draw=none] box {$\ket{0}$} r;
      h r;
      h q;
      phase {$\frac{3\pi}{16}$} q;
      cnot r | q;
      phase {$\frac{-3\pi}{16}$} r;
      cnot r | q;
      phase {$\frac{3\pi}{16}$} r;
      h q;
      measure q;
      %not c[0] | q;
      [operators/every barrier/.append style={-{Latex[length=2.5mm]}, double, double distance=.4mm, line width=.15mm, solid, shorten <= 2.5mm, shorten >= 2.5mm}, operator/separation=1pt]
    barrier (q, r, c[0]);
      box {$c \ll= 1$} (c[0], c[1], c[2]);
      settype {qubit} q;
      [draw=none] box {$\ket{0}$} q;
      h q;
      phase {$\frac{3\pi}{8}$} q;
      cnot r | q;
      phase {$\frac{-3\pi}{8}$} r;
      cnot r | q;
      phase {$\frac{3\pi}{8}$} r;
      phase {$-c$} q;
      h q;
      measure q;
      %not c[0] | q;
      [operators/every barrier/.append style={-{Latex[length=2.5mm]}, double, double distance=.4mm, line width=.15mm, solid, shorten <= 2.5mm, shorten >= 2.5mm}, operator/separation=1pt]
    barrier (q, r, c[0]);
      box {$c \ll= 1$} (c[0], c[1], c[2]);
      settype {qubit} q;
      [draw=none] box {$\ket{0}$} q;
      h q;
      phase {$\frac{3\pi}{4}$} q;
      cnot r | q;
      phase {$\frac{-3\pi}{4}$} r;
      cnot r | q;
      phase {$\frac{3\pi}{4}$} r;
      phase {$-c$} q;
      h q;
      measure q;
      %not c[0] | q;
      [operators/every barrier/.append style={-{Latex[length=2.5mm]}, double, double distance=.4mm, line width=.15mm, solid, shorten <= 2.5mm, shorten >= 2.5mm}, operator/separation=1pt]
    barrier (q, r, c[0]);
      box {$c \ll= 1$} (c[0], c[1], c[2]);
      settype {qubit} q;
   \end{yquant*}
\end{tikzpicture}
}}

\vspace*{-1ex}
\paragraph{Gate simplification (repeat).} We repeat the gate simplification transformations to reduce the gate count. In this example, phase gates commute through the controls of CNOT gates and can be merged with other phase gates by adding the angle arguments.

\qasmfile{examples/example-7.qasm}
\noindent\makebox[\textwidth]{
\scalebox{.7}{
\begin{tikzpicture}
   \begin{yquant*}[operator/separation=0mm]
      qubit {\qasmline{\reg}\!} q;
      qubit {\qasmline{\reg}\!} r;
      cbit {\qasmline{\reg[\idx]} = 0\!} c[3];
      [draw=none] box {$\ket{0}$} q;
      [draw=none] box {$\ket{0}$} r;
      h r;
      h q;
      cnot r | q;
      phase {$\frac{-3\pi}{16}$} r;
      cnot r | q;
      phase {$\frac{3\pi}{16}$} r;
      phase {$\frac{3\pi}{16}$} q;
      h q;
      measure q;
      %not c[0] | q;
      [operators/every barrier/.append style={-{Latex[length=2.5mm]}, double, double distance=.4mm, line width=.15mm, solid, shorten <= 2.5mm, shorten >= 2.5mm}, operator/separation=1pt]
    barrier (q, r, c[0]);
      box {$c \ll= 1$} (c[0], c[1], c[2]);
      settype {qubit} q;
      [draw=none] box {$\ket{0}$} q;
      h q;
      cnot r | q;
      phase {$\frac{-3\pi}{8}$} r;
      cnot r | q;
      phase {$\frac{3\pi}{8}$} r;
      phase {$\frac{3\pi}{8} - c$} q;
      h q;
      measure q;
      %not c[0] | q;
      [operators/every barrier/.append style={-{Latex[length=2.5mm]}, double, double distance=.4mm, line width=.15mm, solid, shorten <= 2.5mm, shorten >= 2.5mm}, operator/separation=1pt]
    barrier (q, r, c[0]);
      box {$c \ll= 1$} (c[0], c[1], c[2]);
      settype {qubit} q;
      [draw=none] box {$\ket{0}$} q;
      h q;
      cnot r | q;
      phase {$\frac{-3\pi}{4}$} r;
      cnot r | q;
      phase {$\frac{3\pi}{4}$} r;
      phase {$\frac{3\pi}{4} - c$} q;
      h q;
      measure q;
      %not c[0] | q;
      [operators/every barrier/.append style={-{Latex[length=2.5mm]}, double, double distance=.4mm, line width=.15mm, solid, shorten <= 2.5mm, shorten >= 2.5mm}, operator/separation=1pt]
    barrier (q, r, c[0]);
      box {$c \ll= 1$} (c[0], c[1], c[2]);
      settype {qubit} q;
   \end{yquant*}
\end{tikzpicture}
}}

\vspace*{-1ex}
\paragraph{Physical qubit mapping and routing.} In this step we need to use the physical connectivity of the target device to map virtual qubits of the circuit to physical qubits of the device. In this example, we assume a CNOT can be applied between physical qubits 0 and 1 in either direction, so no routing is necessary.

\qasmfile{examples/example-9.qasm}
\noindent\makebox[\textwidth]{
\scalebox{.6}{
\begin{tikzpicture}
   \begin{yquant*}[operator/separation=0mm]
      qubit {\texttt{\$0}} q;
      qubit {\texttt{\$1}} r;
      cbit {\qasmline{\reg[\idx]} = 0\!} c[3];
      [draw=none] box {$\ket{0}$} q;
      [draw=none] box {$\ket{0}$} r;
      h r;
      h q;
      cnot r | q;
      phase {$1.8125\pi$} r;
      cnot r | q;
      phase {$0.1875\pi$} r;
      phase {$0.1875\pi$} q;
      h q;
      measure q;
      %not c[0] | q;
      [operators/every barrier/.append style={-{Latex[length=2.5mm]}, double, double distance=.4mm, line width=.15mm, solid, shorten <= 2.5mm, shorten >= 2.5mm}, operator/separation=1pt]
    barrier (q, r, c[0]);
      box {$c \ll= 1$} (c[0], c[1], c[2]);
      settype {qubit} q;
      [draw=none] box {$\ket{0}$} q;
      h q;
      cnot r | q;
      phase {$1.625\pi$} r;
      cnot r | q;
      phase {$0.375\pi$} r;
      phase {$0.375\pi - c$} q;
      h q;
      measure q;
      %not c[0] | q;
      [operators/every barrier/.append style={-{Latex[length=2.5mm]}, double, double distance=.4mm, line width=.15mm, solid, shorten <= 2.5mm, shorten >= 2.5mm}, operator/separation=1pt]
    barrier (q, r, c[0]);
      box {$c \ll= 1$} (c[0], c[1], c[2]);
      settype {qubit} q;
      [draw=none] box {$\ket{0}$} q;
      h q;
      cnot r | q;
      phase {$1.25\pi$} r;
      cnot r | q;
      phase {$0.75\pi$} r;
      phase {$0.75\pi - c$} q;
      h q;
      measure q;
      %not c[0] | q;
      [operators/every barrier/.append style={-{Latex[length=2.5mm]}, double, double distance=.4mm, line width=.15mm, solid, shorten <= 2.5mm, shorten >= 2.5mm}, operator/separation=1pt]
    barrier (q, r, c[0]);
      box {$c \ll= 1$} (c[0], c[1], c[2]);
      settype {qubit} q;
   \end{yquant*}
\end{tikzpicture}
}}

\vspace*{-1ex}
\paragraph{Scheduling.} We can enforce a scheduling policy without knowing or using concrete gate durations. We use stretchy delays for this purpose, which enact a ``as late as possible'' schedule. The actual gate durations will be available in the next stage where the circuit is tied to certain gate calibration data, and the stretch problem can be solved.%

\qasmfile{examples/example-10.qasm}
\noindent\makebox[\textwidth]{
\scalebox{.6}{
\begin{tikzpicture}
  \begin{yquant*}[operator/separation=0mm]
      qubit {\texttt{\$0}} q;
      qubit {\texttt{\$1}} r;
      cbit {\qasmline{\reg[\idx]} = 0\!} c[3];
      [draw=none] box {$\ket{0}$} q;
      [draw=none] box {$\ket{0}$} r;
      h r;
      h q;
      ["s1", name=boxa, x radius=2mm, draw=none] box {} q;
      ["s2", name=boxb, x radius=2mm, draw=none] box {} r;
      cnot r | q;
      phase {$1.8125\pi$} r;
      cnot r | q;
      phase {$0.1875\pi$} r;
      phase {$0.1875\pi$} q;
      h q;
      measure q;
      %not c[0] | q;
      [operators/every barrier/.append style={-{Latex[length=2.5mm]}, double, double distance=.4mm, line width=.15mm, solid, shorten <= 2.5mm, shorten >= 2.5mm}, operator/separation=1pt]
    barrier (q, r, c[0]);
      box {$c \ll= 1$} (c[0], c[1], c[2]);
      settype {qubit} q;

      %[draw=none] box {..} (q, r, c);
      [draw=none] box {$\ket{0}$} q;
      h q;
      ["s3", name=boxc, x radius=2mm, draw=none] box {} r;
      cnot r | q;
      phase {$1.625\pi$} r;
      cnot r | q;
      phase {$0.375\pi$} r;
      phase {$0.375\pi - c$} q;
      h q;
      measure q;
      %not c[0] | q;
      [operators/every barrier/.append style={-{Latex[length=2.5mm]}, double, double distance=.4mm, line width=.15mm, solid, shorten <= 2.5mm, shorten >= 2.5mm}, operator/separation=1pt]
    barrier (q, r, c[0]);
      box {$c \ll= 1$} (c[0], c[1], c[2]);
      settype {qubit} q;

      settype {qubit} q;
      [draw=none] box {$\ket{0}$} q;
      h q;
      ["s4", name=boxd, x radius=2mm, draw=none] box {} r;
      cnot r | q;
      phase {$1.25\pi$} r;
      cnot r | q;
      phase {$0.75\pi$} r;
      phase {$0.75\pi - c$} q;
      h q;
      measure q;
      %not c[0] | q;
      [operators/every barrier/.append style={-{Latex[length=2.5mm]}, double, double distance=.4mm, line width=.15mm, solid, shorten <= 2.5mm, shorten >= 2.5mm}, operator/separation=1pt]
    barrier (q, r, c[0]);
      box {$c \ll= 1$} (c[0], c[1], c[2]);
      settype {qubit} q;
      ["s5", name=boxf, x radius=2mm, draw=none] box {} r;
  \end{yquant*}
  \spring{boxa}{1mm}
  \spring{boxb}{1mm}
  \spring{boxc}{1mm}
  \spring{boxd}{1mm}
  \spring{boxf}{1mm}
\end{tikzpicture}
}}

\vspace*{-1ex}
\paragraph{Calibration linking and stretch resolution.} A pulse sequence is defined for each gate, reset, and measurement. Using their durations, the stretch variables are resolved into delays with concrete timing. Angles are rounded to the precision of the defcal arguments (i.e., to defcal precision) at this step, if they have not already been rounded to the appropriate precision by an earlier transformation.

\qasmfile{examples/example-11.qasm}
\noindent\makebox[\textwidth]{
\scalebox{.65}{
\begin{tikzpicture}
  \begin{yquant*}[operator/separation=0mm]
      qubit {\texttt{\$0}} q;
      qubit {\texttt{\$1}} r;
      cbit {\qasmline{\reg[\idx]}\!} c[3];
      [name=lefttop, x radius=15mm, y radius=3mm] box {$Reset [200]$} q;
      [name=leftbot, x radius=15mm, y radius=3mm] box {$Reset [200]$} r;
      [x radius=6mm] box {$H [20]$} q;
      [x radius=3mm] box {$d [10]$} q;
      [x radius=11.5mm] box {$H [30]$} r;
      [name=cx1, x radius=20mm, y radius=7mm] box {$CX [300]$} (q,r);
      [name=cx2, x radius=20mm, y radius=7mm] box {$CX [300]$} (q,r);
      [x radius=6mm] box {$H [20]$} q;
      [name=righttop, x radius=25mm] box {$Measure [400]$} q;
      [name=rightbot, x radius=31mm] box {$d [420]$} r;
      settype {cbit} q;
      %not c[0] | q;
      [operators/every barrier/.append style={-{Latex[length=2.5mm]}, double, double distance=.4mm, line width=.15mm, solid, shorten <= 2.5mm, shorten >= 2.5mm}, operator/separation=1pt]
    barrier (q, r, c[0]);
      settype {qubit} q;
      box {$c \ll= 1$} (c[0], c[1], c[2]);

      [x radius=15mm, y radius=3mm] box {$Reset [200]$} q;
      [x radius=6mm] box {$H [20]$} q;
      [x radius=21mm] box {$d [220]$} r;

      [draw=none] box {...} (q, r, c);
      discard q;
      discard r;
      discard c;

  \end{yquant*}
  \node[draw, dashed, fit=(lefttop) (leftbot) (righttop) (rightbot)] {};
\circledarrow{gray}{cx1.east}{1.5mm};
\circledarrow{gray}{cx2.east}{1.5mm};
\circledarrow{gray}{cx2.south east}{1.5mm};
\end{tikzpicture}
}}

% cals

% Phase defcal

\begin{tikzpicture}[baseline=(current bounding box.center)]
  \begin{yquant*}
      qubit q;
      hspace {2mm} q;
      phase {$\theta$} q;
      hspace {2mm} q;
  \end{yquant*}
\end{tikzpicture}
$=$
\begin{tikzpicture}[baseline=(current bounding box.center)]
  \begin{yquant*}
      qubit {$q.drive$} d;
      [name=box, draw=none, x radius=1mm, y radius=1mm]
      box {} d;
  \end{yquant*}
\circledarrow{gray}{box.west}{1.5mm};
\draw (box.west) -- (box.east);
\end{tikzpicture}

\vspace{5mm}

% H0 defcal
\begin{tikzpicture}[baseline=(current bounding box.center)]
  \begin{yquant*}
      qubit {$\$0$} q;
      h q;
  \end{yquant*}
\end{tikzpicture}
$=$
\raisebox{1mm}{
\begin{tikzpicture}[baseline=(current bounding box.center)]
  \begin{yquant*}
      qubit {$\$0.drive$} d;
      [name=box, draw=none, x radius=5mm, y radius=5mm]
      box {} d;
  \end{yquant*}
\def\gaussian{\x,{4*1/exp(((\x-3)^2)/2)}}
\begin{scope}[x=2mm,y=2mm,shift={(0mm,-5mm)}]
\circledarrow{gray}{box.west}{1.5mm};
\draw[thick] (0,0) plot[domain=0:6] (\gaussian);
\circledarrow{gray}{box.east}{1.5mm};
\draw (box.west) -- (box.east);
\end{scope}
\end{tikzpicture}
}

\vspace{5mm}

% H1 defcal
\begin{tikzpicture}[baseline=(current bounding box.center)]
  \begin{yquant*}
      qubit {$\$1$} q;
      h q;
  \end{yquant*}
\end{tikzpicture}
$=$
\begin{tikzpicture}[baseline=(current bounding box.center)]
  \begin{yquant*}
      qubit {$\$1.drive$} d;
      [name=box, draw=none, x radius=8.5mm, y radius=5mm]
      box {} d;
  \end{yquant*}
\def\gaussian{\x,{2*1/exp(((\x-5)^2)/5)}}
\begin{scope}[x=2mm,y=2mm,shift={(0mm,-5mm)}]
\circledarrow{gray}{box.west}{1.5mm};
\draw[thick] (0,0) plot[domain=0:10] (\gaussian);
\circledarrow{gray}{box.east}{1.5mm};
\draw (box.west) -- (box.east);
\end{scope}
\end{tikzpicture}

\vspace{3mm}

% CX defcal
\begin{tikzpicture}[baseline=(current bounding box.center)]
  \begin{yquant*}
      qubit {$\$0$} ctrl;
      qubit {$\$1$} tgt;
      cnot tgt | ctrl;
  \end{yquant*}
\end{tikzpicture}
$=$
\raisebox{4mm}{
\begin{tikzpicture}[baseline=(current bounding box.center)]
  \begin{yquant*}[operator/separation=0mm]
      qubit {$\$0.drive$} ctrl;
      qubit {$\$1.drive$} tgt;
      hspace {5mm} ctrl;
      [name=box1, draw=none, x radius=5mm, y radius=5mm] box {} ctrl;
      [name=box5, draw=none, x radius=7.5mm, y radius=5mm] box {} tgt;
      barrier (ctrl, tgt);
      [name=box2, draw=none, x radius=12mm, y radius=5mm] box {} ctrl;
      barrier (ctrl, tgt);
      [name=box3, draw=none, x radius=4mm, y radius=5mm] box {} ctrl;
      barrier (ctrl, tgt);
      [name=box4, draw=none, x radius=12mm, y radius=5mm] box {} ctrl;
      barrier (ctrl, tgt);
  \end{yquant*}
\def\gaussiantall{\x,{6*1/exp(((\x-3)^2)/2)}}
\def\gaussianshort{\x,{2*1/exp(((\x-4)^2)/5)}}
\def\gaussianflat{\x,{3*1/exp(((\x-4)^2)/2)}}
\def\flat{\x,{3)}}

\begin{scope}[x=2mm,y=2mm,shift={(5mm,-5mm)}]
\draw[thick] (0,0) plot[domain=0:6] (\gaussiantall);
\end{scope}

\begin{scope}[x=2mm,y=2mm,shift={(15mm,-5mm)}]
\draw[thick] (0,0) plot[domain=1:4] (\gaussianflat);
\end{scope}

\begin{scope}[x=2mm,y=2mm,shift={(22.5mm,-5mm)}]
\draw[thick] (0,0) plot[domain=0:9.5] (\flat);
\end{scope}

\begin{scope}[x=2mm,y=2mm,shift={(33mm,-5mm)}]
\draw[thick] (0,0) plot[domain=4:7] (\gaussianflat);
\end{scope}

\begin{scope}[x=2mm,y=2mm,shift={(47mm,-5mm)}]
\draw[thick] (0,0) plot[domain=0:6] (\gaussiantall);
\end{scope}

\begin{scope}[x=2mm,y=2mm,shift={(57mm,-5mm)}]
\draw[thick] (0,0) plot[domain=1:4] (\gaussianflat);
\end{scope}

\begin{scope}[x=2mm,y=2mm,shift={(64mm,-5mm)}]
\draw[thick] (0,0) plot[domain=0:9.5] (\flat);
\end{scope}

\begin{scope}[x=2mm,y=2mm,shift={(75mm,-5mm)}]
\draw[thick] (0,0) plot[domain=4:7] (\gaussianflat);
\end{scope}

\begin{scope}[x=2mm,y=2mm,shift={(0mm,-16mm)}]
\draw[thick] (0,0) plot[domain=0:9] (\gaussianshort);
\end{scope}

\circledarrow{gray}{box1.west}{1.5mm};
\circledarrow{gray}{box4.west}{1.5mm};
\circledarrow{gray}{box4.east}{1.5mm};
\draw (box1.west) -- (box1.east);
\draw (box2.west) -- (box2.east);
\draw (box3.west) -- (box3.east);
\draw (box4.west) -- (box4.east);
\draw (box5.west) -- (box5.east);
\end{tikzpicture}
}

\vspace{4mm}

% Meas defcal
\begin{tikzpicture}[baseline=(current bounding box.center)]
  \begin{yquant*}
      qubit {$\$0$} q;
      measure q;
  \end{yquant*}
\end{tikzpicture}
$=$
\raisebox{2mm}{
\begin{tikzpicture}[baseline=(current bounding box.center)]
  \begin{yquant*}
      qubit {$\$0.measure$} meas;
      qubit {$\$0.acquire$} acq;
      [name=box1, draw=none, x radius=50mm, y radius=5mm] box {} meas;
      [name=box2, draw=none, x radius=50mm, y radius=5mm] box {} acq;
  \end{yquant*}

\def\gaussianflat{\x,{3*1/exp(((\x-4)^2)/2)}}
\def\flat{\x,{3)}}

\begin{scope}[x=2mm,y=2mm,shift={(0mm,-5mm)}]
\draw[thick] (0,0) plot[domain=1:4] (\gaussianflat);
\end{scope}

\begin{scope}[x=2mm,y=2mm,shift={(7.5mm,-5mm)}]
\draw[thick] (0,0) plot[domain=0:43] (\flat);
\end{scope}

\begin{scope}[x=2mm,y=2mm,shift={(84.5mm,-5mm)}]
\draw[thick] (0,0) plot[domain=4:7] (\gaussianflat);
\end{scope}

\begin{scope}[x=2mm,y=2mm,shift={(1mm,-20mm)}]
\draw[thick] (0,0) plot[domain=0:50] (\flat);
\end{scope}

\draw (box1.west) -- (box1.east);
\draw (box2.west) -- (box2.east);
\end{tikzpicture}
}

The compiled circuit is now submitted to the target machine code generator to produce binaries for the target control system of the quantum computer. These are then submitted to the execution engine to orchestrate the quantum computation. Additional phases of target-dependent compilation occur that are beyond the scope of OpenQASM 3.

\section{Related Work and Future Extensions}

The design of OpenQASM\,3 was naturally influenced by features present in other quantum programming languages, as well as constructions which are standard concepts in the quantum computing literature and which were considered feasible to include.
A number of other features or design choices were also considered.
In some cases, these were set aside in the interests of some other design choice; in other cases they were considered to be possibly worth pursuing but not a core functionality.
In this Section, we describe how some of the `new' features in OpenQASM\,3 (\emph{i.e.},~which extend what existed in OpenQASM\,2) relate to the features of other quantum programming languages, which choices were considered but not included in OpenQASM\,3, and which choices are still under consideration for inclusion.

\subsection{Comparison to other quantum programming languages}

There has been significant prior work on quantum programming languages and intermediate representations, some of which have directly influenced OpenQASM.
Many high-level quantum programming languages are embedded in classical languages. For example, Scaffold~\cite{scaffold}, QCL~\cite{omer03}, and qcor~\cite{qcor21}  are embedded in C or C\texttt{++}, Quipper~\cite{quipper} in Haskell, ProjectQ~\cite{projectq} in Python, and QIR~\cite{QIR} in LLVM. On the other hand languages such as OpenQASM and Quil~\cite{smith16} are designed to be standalone and portable intermediate representations.

Classical control flow exists in many of the above languages.
These are sometimes only used as a means of code compression (\emph{i.e.},~fully unrollable at compile time)~\cite{projectq,scaffold}, and sometimes are intended to mix classical and quantum compute~\cite{smith16,qsharp}.
In embedded languages the host language provides direct access to classical types and control flow. OpenQASM\,3 expands upon these by introducing new types such as \qasmline{angle}, and allowing the semantic definition of purely classical \qasmline{extern} functions in the context of real-time computation.

Gate modifiers (at least for ``control''), exist in most of the above programming languages.
In OpenQASM\,3 we utilize these modifiers for two reasons.
First, they decouple semantics from implementation --- many implementations could exist for a particular modified gate.
Second, modifiers in conjunction with two simple language built-ins (\qasmline{U} gate and \qasmline{gphase}), generate the entire standard gate library.

The use of timing and pulse-level control within quantum circuits has long been present in  software and languages used for controlling experiments.
Rigorous specifications for this were introduced for example in QGL~\cite{QGL}, OpenPulse~\cite{mckay2018}, eQASM~\cite{eqasm19}, Artiq~\cite{artiq}, Pulser~\cite{pulser}, Jaqal~\cite{jacal}, Qua~\cite{qua}, and Quil-t~\cite{quilt}.
OpenQASM\,3 takes this further by streamlining the incorporation of timing in the circuit model, by allowing timing constraints to be expressed at a level that is decoupled from actual pulse implementations (using \qasmline{stretch}).
This further motivated the inclusion of a scoping mechanism in OpenQASM\,3 in the form of \qasmline{box}, which allows references to specific blocks of code and to guide compiler optimizations.
This has been addressed before in a more limited form using barriers~\cite{CBSG17} and pragmas~\cite{smith16}.

\subsection{Features considered but not adopted}

In this article we have presented the design decisions in OpenQASM\,3 and the rationale behind them. In a similar vein, multiple other language features were contemplated but abandoned or left for future revisions of the language. In the following we briefly touch upon these.

Some languages such as Q\# and ProjectQ support resource and memory management, in the form of mid-circuit allocation and de-allocation of qubits. Some languages such as Scaffold allow for the expression of classical code that will then be compiled to reversible quantum circuits (e.g. oracles). In each of these cases, OpenQASM intentionally remains lower-level and more grounded in executable circuits, aiming to keep the compiler's complexity more manageable.

Conversely, some languages such as Quil have unstructured control flow in the form of explicit program labels and {\tt JUMP} statements, which are in fact more in the style of a classical assembly language.
In order to retain backwards compatibility with OpenQASM\,2, we chose not to require any \emph{explicit} entry points, preferring to allow the program to proceed from the first instruction in a global scope.
OpenQASM\,3 also adopted the use of {\tt for} and {\tt while} loops to express control flow, both to retain easier programmability by humans, and to allow for easier timing analysis by the compiler.
These all motivated a design philosophy that it should not be necessary in OpenQASM\,3 to manage the (classical) control flow, on a similarly low level to management of quantum operations and resources.

\subsection{Considering further language features for OpenQASM}

The OpenQASM language will continue to evolve however within its governance model, and certain features may be added.
An example of such a feature are unitary circuit families and generic subroutines, corresponding roughly to templated function definitions, which may be defined entirely in OpenQASM\,3.
(While such subroutines could be written in a separate host language, writing them in OpenQASM\,3 allows for greater portability.)
Another language feature which is at an advanced stage of consideration at the time of writing it the inclusion of classical arrays as a container type; a third feature which has been set aside, but has been re-proposed for consideration is a fixed-point real number type.
The usefulness of such features, the way that they interoperate or conflict with other features, and how they may be incorporated into the language and supported as features is the subject of discussion in working groups established for this purpose.
In this way, OpenQASM\,3 will continue to evolve, incorporating features which are sufficiently important and consistent with the other functionality which it provides, through input from the community.

\section{Conclusion}

OpenQASM\,3 has introduced language features that both expand and deepen the scope quantum circuits that can be described with particular focus on their physical implementation and the interactions between classical and quantum computing. We have endeavored throughout to give a consistent ``look and feel'' to these features so that regardless of abstraction level, OpenQASM\,3 still feels like one language. This manuscript has illustrated the primary new features introduced in OpenQASM\,3 and put those features in context with examples. In particular, we have shown features and representations appropriate to a variety of abstraction levels and use cases, from high-level gate modifiers to low-level microcoded gate implementations.

Language design is an open-ended problem and we expect that as we attempt to use OpenQASM\,3 to program circuits on our primitive quantum computers we will discover many awkward constructions, missing features, or incomplete specifications in the language. We fully expect the language to continue to evolve over time driven by real world usage and hardware development. In particular, there are already proposals under consideration to modify implicit type conversions or enable code re-use through generic functions. Consequently, the formal language specification is posted as a live document~\cite{livedoc} within the Qiskit project. We will be forming a formal process for reviewing proposed changes so that OpenQASM might continue to adapt to the needs of its users.

\section{Acknowledgements}

We thank the community who gave early feedback on the OpenQASM 3 spec~\cite{livedoc}, especially Hossein Ajallooiean, Luciano Bello, Yudong Cao, Lauren Capelluto, Pranav Gokhale, Michael Healey, Hiroshi Horii, Sonika Johri, Peter Karalekas, Moritz Kirste, Kevin Krsulich, Andrew Landahl, Prakash Murali, Salva de la Puente, Kenneth Rudinger, Zachary Schoenfeld, Yunong Shi, Robert Smith, Stefan Teleman, Ntwali Bashinge Toussaint, and Jack Woehr.
NdB's contributions build on work with Jacob Wilkins, as part of the EPSRC-funded NQIT project and as the Oxford--Tencent Fellow in the Department of Computer Science at Oxford.
Quantum circuit drawings were typeset using yquant~\cite{desef2020yquant}.

\bibliographystyle{unsrt}
\bibliography{qasm3}

\end{document}